\documentclass[10pt,showkeys,twocolumn,superscriptaddress]{revtex4}
\usepackage[draft]{hyperref}
\usepackage{amssymb,amsmath}
\usepackage{multirow}
\usepackage{array}
\usepackage[T1]{fontenc}
\usepackage{textcomp}
\usepackage{ae}
\usepackage{soul} % for strikethrough
\usepackage{graphicx}

\makeatletter

\newcommand\subnumber[1]{\textbf{\sffamily #1}}
\long\def\@makecaption#1#2{%
  \vskip\abovecaptionskip
  \sbox\@tempboxa{\captionfont\textbf{\sffamily #1}\quad #2}%
  \ifdim \wd\@tempboxa >\hsize
    {\captionfont\textbf{\sffamily #1}\quad #2\par}%
  \else
    \global \@minipagefalse
    \hb@xt@\hsize{\hfil\box\@tempboxa\hfil}%
  \fi
  \vskip\belowcaptionskip}

\newcommand\floatfont{\small}
\newcommand\captionfont{\small}

\renewcommand*\refstepcounter[1]{\stepcounter{#1}%
  \protected@edef\@currentlabel{%
    \csname p@#1\expandafter\endcsname
    \csname the#1\endcsname
  }%
}

\newcounter{circulator}
\newcommand\declarecirculator[1]{\refstepcounter{circulator}\label{#1}}
\renewcommand\p@circulator[1]{C#1}
\newcounter{geometry}
\newcommand\declaregeometry[1]{\refstepcounter{geometry}\label{#1}}
\renewcommand\p@geometry[1]{G#1}

\makeatother

\DeclareMathOperator{\E}{e}

\newcommand{\vect}[1]{\boldsymbol{#1}}

\newcommand{\mat}[1]{\hat{#1}}
\newcommand{\vers}[1]{\boldsymbol{\hat{#1}}}
\newcommand{\I}{\mathrm{i}}
\newcommand{\diff}{\mathrm{d}}
\newcommand{\transpose}{^\mathrm{T}}

\newcommand{\pderiv}[3][]{\frac{\partial^{#1} #2}{\partial #3^{#1}}}
\newcommand{\tsub}[1]{_{\text{#1}}}

\newcommand{\abs}[1]{\lvert #1 \rvert}

\usepackage{booktabs}

\newcommand{\phz}{\hphantom{0}}

\graphicspath{{Figures/}}
\begin{document}

\title{Compact optical circulator based on a uniformly magnetized ring cavity}
\author{Wojciech \'Smigaj}
\altaffiliation[Currently with ]{Laboratoire Charles Fabry de l'Institut d'Optique, CNRS, Universit\'e Paris XI, Palaiseau, France}
\email{wojciech.smigaj@institutoptique.fr}
\affiliation{Institut Fresnel, CNRS UMR6133, Universit\'e Aix--Marseille III, Marseille, France}
\author{Liubov Magdenko}
\altaffiliation[Currently with ]{Laboratoire de Photonique et de Nanostructures, CNRS UPR20, Marcoussis, France}
\affiliation{Institut d'\'Electronique Fondamentale, Universit\'e Paris XI, Orsay, France}
\author{Javier Romero-Vivas}
\altaffiliation[Currently with ]{Surface Physics Division, Faculty of Physics, Adam Mickiewicz University, ul.\ Umultowska 85, 61-614 Pozna\'n, Poland}
\affiliation{Institut Fresnel, CNRS UMR6133, Universit\'e Aix--Marseille III, Marseille, France}
\author{S\'ebastien Guenneau}
\affiliation{Institut Fresnel, CNRS UMR6133, Universit\'e Aix--Marseille III, Marseille, France}
\author{B\'eatrice Dagens}
\affiliation{Institut d'\'Electronique Fondamentale, Universit\'e Paris XI, Orsay, France}
\author{Boris Gralak}
\affiliation{Institut Fresnel, CNRS UMR6133, Universit\'e Aix--Marseille III, Marseille, France}
\author{Mathias Vanwolleghem}
\affiliation{Institut d'\'Electronique Fondamentale, Universit\'e Paris XI, Orsay, France}

\begin{abstract} 
We propose a new class of compact integrated optical circulators providing a large isolation level while maintaining a straightforward technological feasibility. Their layout is based on a nonreciprocal radial Bragg cavity composed of concentric magneto-optical rings. The circulator ports are standard rib waveguides, butt-coupled to the cavity by cutting through its outer rings. The device is specifically designed for operation in a uniform external magnetic field. Using a coupled-mode description of the complete cavity/waveguide-port system, we explore the rich behaviour of cavity circulators in presence of varying levels of direct port-to-port coupling. We demonstrate numerically a strongly miniaturized two-dimensional cavity circulator, with a total footprint of less than $(10\lambda)^2$, achieving a 20-dB isolation level at telecom frequencies over a bandwidth of 130\,GHz. The device is found to be very tolerant with respect to fabrication imperfections. We finish with an outlook on three-dimensional versions of the proposed nonreciprocal cavities.
\end{abstract}


\maketitle

\section{Introduction}\label{sec:Intro}
The need for an integrated and miniaturized version of an optical isolator, or more generally an optical circulator, is making itself increasingly felt. The drive towards ever higher degrees of all-optical on-chip integration is often hindered by the absence of an element that induces a one-way sense in the path of the signal in such an integrated circuit. Without it, long path interferences can lead to large changes somewhere in the circuit due to small amplitude oscillations at a remote point. Commercially available isolators are bulk free-space devices based on 45$^\circ$ nonreciprocal magneto-optical (MO) Faraday polarization rotators in combination with polarizers placed at their entrance and exit. The most commonly used MO materials are magnetic garnet oxides such as Ce-substituted Ce$_x$Y$_{3-x}$Fe$_5$O$_{12}$ (Ce:YIG), combining optical transparency and strong MO properties at telecom frequencies. Realisation of an integrated isolator based on the Faraday-effect is very difficult because of the inevitable geometric birefringence of planar integrated waveguide circuits.

Nowadays, research efforts focus on integrated isolator concepts that do not depend on polarization conversion~\cite{DotschJOSAB05}. This can be achieved by properly orienting the magnetization in the MO material so that no coupling occurs between the quasi-TE and quasi-TM waveguide modes. Phase velocity and field profiles of the waveguide modes then become different for forward and backward propagation, while their polarization state is unchanged. This has been exploited to propose various concepts based on nonreciprocal interference~\cite{AurOptCom75,ShoAPL08}, multimode imaging~\cite{ZhuOQE00}, microring and -disk resonators \cite{KonoOE07,JalasOL2010}, etc. Nevertheless, experimental demonstration of garnet-based isolators suffers from the limited MO gyrotropy in the optical and near-infrared regime. This leads to high device lengths (typically of the order of 1\,mm), which goes hand in hand with serious technological challenges, for instance to maintain magnetic uniformity. As a result, during the past decade the use of resonant photonic-crystal (PhC) layouts has attracted a lot of interest. Magnetophotonic crystals allow to artificially enhance the intrinsic strength of the basic MO effects \cite{InoueJPD06,SteelJLT00}; in addition, removal of time-reversal symmetry in periodic structures leads to the appearance of novel phenomena, such as frozen light~\cite{FigotinPRE01}, photonic chiral edge states~\cite{HaldanePRL08} and one-way band gaps~\cite{VanwolleghemPRB09,YuAPL07}. Over the last few years, an increasing number of promising miniaturized isolator (and circulator) designs using PhC effects have been reported~\cite{WangOL05,SmiOL10,VanwolleghemPRB09,KonoJLT2004,TakedaPRA08}.

In this article, we will focus on a particular class of miniaturized integrated circulator designs based on the resonant enhancement of the light--MO-material interaction in a nonreciprocal cavity. Uniformly magnetized resonant ferrite cavities have been used for decades in microwave circulators~\cite{Pozarbookch9}. However, optical circulators made by a simple geometric rescaling of existing microwave devices would have prohibitively low operation bandwidths. This is because at optical frequencies nonreciprocal effects are induced by the gyroelectric off-diagonal elements of the permittivity tensor, which are typically one or two orders of magnitude smaller than the analogous gyromagnetic off-diagonal elements of the permeability tensor in the microwave regime~\cite{Visnovskybook}. In 2005, Wang and Fan reported a solution for a cavity circulator operating at optical frequencies~\cite{WangOL05}. The proposed device is composed of a 2D PhC cavity etched in bismuth iron garnet (BIG) --- a transparent magnetic oxide with record MO properties~\cite{VertruyenPRB08}. The cavity is evanescently coupled to three symmetrically placed PhC waveguides. It achieves an infinitely strong circulation at the resonant wavelength of the cavity and has a footprint of just a few square wavelengths. However, under uniform magnetization the spectral bandwidth of this circulator becomes negligibly small, reducing ultimately the applicability of the device. A reasonable bandwidth of the order of 50\,GHz can only be achieved by imposing a very specific domain structure of antiparallel magnetic domains. Achieving and maintaining this magnetic domain structure within an area of a few \textmu m$^2$ is unfeasible.

\begin{figure}[!tb]
\centering
  \includegraphics{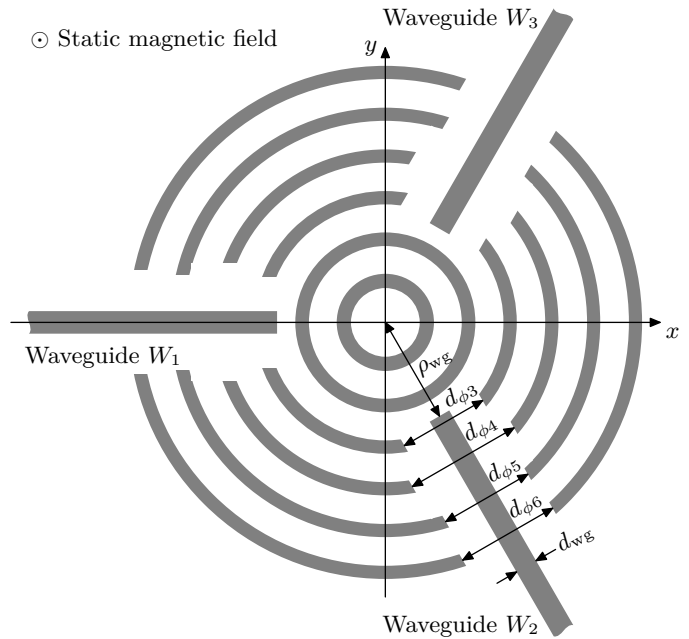}
\caption{Geometry of a circulator composed of a ring cavity butt-coupled to three rib-waveguides. A uniform magnetic field perpendicular to the $xy$~plane causes nonreciprocal coupling of the even and odd degenerate $H$-polarized cavity modes. The number of split and full rings, $n\tsub{s}$ resp.\ $n\tsub{f}$, the width of the slits in the split rings, $d_{\phi n}$, and the distance from the waveguide ends to the centre of the cavity, $\rho\tsub{wg}$, are all design optimization parameters.}
\label{fig:geometry-general}
\end{figure}

In order to remedy the unfeasibility of this concept we reported earlier an original design approach for a PhC cavity that provides simultaneously high circulation levels ($\geq 30$\,dB), good spectral bandwidth ($\sim$80\,GHz), and operates in a uniformly magnetized MO material~\cite{SmiOL10}. The new concept is based on an approximately axisymmetric arrangement of the holes making up the PhC cavity. In this work we propose to eliminate entirely the use of the PhC layout, forming instead a fully axisymmetric cavity composed of concentric MO rings and butt-coupling it to standard rib waveguides. The general geometry of the proposed new cavity circulator is shown in Fig.~\ref{fig:geometry-general}. If this device shows similar performance as the cavity circulators proposed in Refs.~\onlinecite{WangOL05,SmiOL10} and~\onlinecite{KonoOE07}, its markedly simpler layout would make it largely preferable over the existing circulator schemes. There are nevertheless important issues to be tackled. The circular Bragg grating formed by the concentric full and split rings can in theory provide sufficient confinement~\cite{SchonenbergerOE09,ScheuerJOSAB07}. However, the quality factor will be lower than that of a PhC cavity. Moreover, this ring circulator is not embedded in a PhC crystal and thus does not operate in a bandgap. As a result, its operation can be perturbed by power loss to free space and direct waveguide-to-waveguide coupling. This latter mechanism allows the appearance of Fano resonances in the transmission spectrum of the cavity. In this paper we will demonstrate that, even in the presence of these effects, the circulator layout of Fig.~\ref{fig:geometry-general} can be designed to achieve competitive performance.

This paper is organized as follows. We begin by extending, in Section~\ref{sec:temp_cp_mode}, the temporal coupled-mode model introduced by Wang and Fan~\cite{WangOL05} to the case of non-negligible direct waveguide-to-waveguide coupling and in-plane radiation losses. This will allow a first assessment of the theoretical maximally achievable device performance. Section~\ref{sec:circulator_design} will then focus on the design and performance of the new type of cavity circulator. We will first detail how the ring cavity itself is optimally designed to allow operation in a uniform magnetic field (Section \ref{ssec:cavity_design}), and then deal with the design optimization of the complete circulator (Sections \ref{ssec:circulator-design-intro} and \ref{ssec:circulator-design-optimization}), using the insight provided by the coupled-mode model of Section~\ref{sec:temp_cp_mode}. In this way it will be shown how this type of circulator competes with the PhC cavity layouts of our earlier work~\cite{SmiOL10}. Before concluding, in Section~\ref{sec:perspectives} we will indicate the additional constraints placed on the cavity design by the presence of a super- and substrate, and show initial designs of properly behaving realistic 3D versions of circulator cavities.

%%%%%%%%%%%%%%%%%%%%%%%%%%%%%%%%%%%%%%%%%%%%%%%%%%%%%%%%%%%%%%%%%%%%%%%%%%%%%%%%%%%%%%%%%%%%%%%%%%%%%%%%%%%%%%%%%%%%%%%%%%%%%%%%%%%%%%%%%%%%%%%
\section{Temporal coupled-mode theory of ring-cavity circulators}\label{sec:temp_cp_mode}

In contrast to PhC-based circulators, where the bandgap provides an almost perfect discoupling of the waveguides, in systems based on rib waveguides one cannot \textit{a priori} exclude the possibility of direct waveguide-to-waveguide coupling. In this section we will firstly extend the abstract temporal coupled-wave description of a nonreciprocal resonator to include such direct transmissions (Section \ref{ssec:tcm_lossless}), using the work of \citet{SuhJQE04} as a basis. Once the model equations have been derived, their formal solution, in Section \ref{ssec:solution_tcm}, will allow to identify the critical device parameters and to evaluate the possible detrimental influence of the direct pathway coupling on the circulator performance. As will be shown in Section~\ref{ssec:discussion_tcm}, nonreciprocal Fano-type resonances appear in the circulator transfer function when the device's ports are not perfectly decoupled. This does not necessarily undermine the device's behaviour, however, but even allows novel functionalities. In the last subsection, \ref{ssec:tcm_radiation}, radiative cavity decay will be included in the model and its influence on the circulator performance will be quantitatively evaluated.

%%%%%%%%%%%%%%%%%%%%%%%%%%%%%%%%%%%%%%%%%%%%%%%%%%%%%%%%%%%%%%%%%%%%%%%%%%%%%%%%%%%%%%%%%%%%%%%%%%%%%%%%%%%%%%%%%%%%%%%%%%%%%%%%%%%%%%%%%%%%%%

\subsection{Inclusion of direct pathway coupling}\label{ssec:tcm_lossless}
Let us consider a circulator composed of three identical single-mode waveguides, $W_1$, $W_2$, and~$W_3$,  weakly coupled with a resonant cavity and arranged so that the whole system has $C_{3v}$ symmetry. A particular example of such a system is shown in Fig.~\ref{fig:geometry-general}. In the absence of MO coupling, the cavity is assumed to support a pair of degenerate orthonormal eigenmodes of frequency~$\omega_0$, belonging to the unique two-dimensional irreducible representation of the $C_{3v}$ point group. They can be classified as even or odd, according to their symmetry with respect to reflection about the axis of waveguide~$W_1$, which is assumed to lie along the $-x$~axis. The circulator's operation at frequency~$\omega$ is described by the coupled-mode equations~\cite{SuhJQE04}
\begin{subequations}
  \label{eq:suh-coupled-mode}
  \begin{align}
  \label{eq:suh-coupled-mode-a}
    -\I\omega\vect a &= 
    -(\I\mat\Omega + \mat\Gamma) \vect a +
    \mat D\transpose \vect s\tsub{in},\\
  \label{eq:suh-coupled-mode-b}
    \vect s\tsub{out} &= \mat C \vect s\tsub{in} + \mat D \vect a,
  \end{align}
\end{subequations}
where the vector $\vect a = (a\tsub e, a\tsub o)\transpose$ contains the amplitudes of the even and odd cavity mode, the vectors $\vect s\tsub{in} = (s_{1,\mathrm{in}}, s_{2,\mathrm{in}}, s_{3,\mathrm{in}})\transpose$ and $\vect s\tsub{out} = (s_{1,\mathrm{out}}, s_{2,\mathrm{out}}, s_{3,\mathrm{out}})\transpose$ contain the amplitudes of the in- and outgoing waveguide modes, the matrix $\mat C$ describes the direct coupling between waveguides, $\mat D$ the cavity-waveguide coupling, $\mat \Omega$ the cavity mode eigenfrequencies and their coupling, and $\mat \Gamma$ their decay. These matrices are subject to the fundamental constraints~\cite{SuhJQE04}
\begin{subequations}
  \begin{align}
    \label{eq:suh-constraint-1}
    \mat D^\dagger \mat D &= 2\mat \Gamma,\\
    \label{eq:suh-constraint-2}
    \mat C \mat D^* &= -\mat D,
  \end{align}
\end{subequations}
where $^*$ denotes complex conjugation and $^\dagger$ Hermitian conjugation. Equation~\eqref{eq:suh-constraint-1} means in particular that all the energy of the decaying cavity modes is passed to the waveguides, i.e., the system is lossless as a whole. The case with radiation loss will be investigated later. Together with those following from the system's symmetry, the above constraints can be used to reduce the number of independent parameters necessary for the characterization of the device. 

The time-reversal symmetry breaking of the device is assumed to be entirely described by the cross-coupling of the cavity resonances via the off-diagonal elements of the Hermitian matrix~$\mat\Omega$. In Section~\ref{ssec:cavity_design} we show that these are induced by magnetization perpendicular to the cavity's plane, which couples nonreciprocally its $p$-polarized modes (with the electric field oriented in-plane). The coupling to the waveguide output ports is considered to be reciprocal, as testified by the use of the same cavity-waveguide coupling matrix $\hat{D}$ in both Eqs.~\eqref{eq:suh-coupled-mode-a} and \eqref{eq:suh-coupled-mode-b}. Similarly, the constraint described by Eq.~\eqref{eq:suh-constraint-2} presupposes time-reversal symmetry of the cavity-waveguide coupling~\cite{SuhJQE04}. This ansatz is justified since the main MO interaction in the device is the nonreciprocal coupling of the cavity resonances among themselves. We shall now consider each of the matrices occurring in Eq.~\eqref{eq:suh-coupled-mode} in turn.

Owing to the three-fold rotational symmetry of the circulator, the $\mat C$ matrix must have the form
\begin{equation}
  \label{eq:C-matrix}
  \mat C = \begin{bmatrix} r&t&t\\ t&r&t\\ t&t&r \end{bmatrix},
\end{equation}
where $t \equiv \abs t \E^{\I\tau}$ and $r \equiv \abs r \E^{\I(\pi+\tau+\Delta)}$ are the transmission and reflection coefficients of the waveguide modes in the absence of the cavity. For future convenience we include an explicit term $\pi$ in the phase factor of~$r$. All nonreciprocity is supposed to be contained in the cavity mode coupling, hence the symmetry of $\mat C$. If the system is lossless, as we assume in the present subsection, we can use the condition of unitarity of~$\mat C$ (or $\abs r^2 + 2 \abs t^2 = 1$ and $\abs t \mathbin{-} 2\abs r \cos\Delta = 0$) to express $t$ and~$r$ as
\begin{equation}
  \label{eq:r-and-t}
  t = \frac{2\cos\Delta \E^{\I\tau}}{\sqrt{1 + 8\cos^2\Delta}},
  \qquad
  r = -\frac{\E^{\I(\tau+\Delta)}}{\sqrt{1 + 8\cos^2\Delta}}
\end{equation}
with the additional condition $\cos\Delta \mathrel{\geq} 0$. Without loss of generality we can therefore restrict $\Delta$ to the interval $[-\frac12\pi, \frac12\pi]$. Note that the case of no direct coupling, considered in Ref.~\onlinecite{WangAPB05}, corresponds to $\Delta = \pm \frac12\pi$ (in other words, the direct waveguide transmission and reflection coefficients being in quadrature), while maximum direct coupling occurs when $\Delta=0$ ($r$ and~$t$ in antiphase).

The $\mat D$ matrix has the general form
\begin{equation}
  \label{eq:D-matrix}
  \mat D = 
  \begin{bmatrix} 
    d\tsub{1e} & d\tsub{1o} \\
    d\tsub{2e} & d\tsub{2o} \\
    d\tsub{3e} & d\tsub{3o}
  \end{bmatrix},
\end{equation}
where $d_{im}$ ($i = 1, 2, 3$; $m = \mathrm{e}, \mathrm{o}$) describes the coupling of $m$th cavity mode with $i$th waveguide. These coupling parameters are proportional to the values of the electromagnetic field of the modes along the waveguide axes. Owing to the assumed symmetry of the mode fields and the symmetrical arrangement of the three waveguide ports, these 6 complex parameters can be expressed in terms of a single complex coupling constant $d \equiv \abs d \E^{i\delta}$:
\begin{equation}
  \label{eq:D-matrix-reduced}
  \mat D = d
  \begin{bmatrix} 
    1 & 0\\
    -\frac12 & \frac{\sqrt{3}}{2} \\ 
    -\frac12 & -\frac{\sqrt{3}}{2}
  \end{bmatrix} \equiv d \mat{\tilde{D}}.
\end{equation}
Substituting this formula into Eq.~(\ref{eq:suh-constraint-1}), we obtain 
\begin{equation}
  \label{eq:Gamma-matrix}
  \mat \Gamma = \gamma
  \begin{bmatrix} 1 & 0 \\ 0 & 1 \end{bmatrix},
  \quad\text{where}\quad \gamma\equiv\tfrac 34 \abs d^2  .
\end{equation}
The second constraint, Eq.~(\ref{eq:suh-constraint-2}), yields
\begin{equation}
  \label{eq:intermediate-constraint-on-d}
  (t-r) d^* = d, \qquad \mathrm{or} \qquad d^2 = (t-r) \abs d^2 = \tfrac{4}{3}\gamma (t-r).
\end{equation}
The latter form is important, since we shall see further down that the behaviour of the device depends solely on~$d^2$. This expression therefore shows that the phase $\delta$ of the coupling constant $d$ plays no role. From Eq.~(\ref{eq:r-and-t}) we have 
\begin{equation}
  \label{eq:t-r}
  \begin{split}
  t-r &= 
  \frac{2\cos\Delta \mathbin{+}\E^{\I\Delta}}{\sqrt{1 + 8\cos^2\Delta}}\E^{\I\tau} =
  \frac{3\cos\Delta + \I\sin\Delta}{\abs{3\cos\Delta + \I\sin\Delta}}
  \E^{\I\tau} \\&=
  \E^{\I[\tau + \arg(3\cos\Delta + \I\sin\Delta)]}.
\end{split}
\end{equation}

The coupling between the cavity modes is described by the off-diagonal elements in $\mat\Omega$. Since we presuppose nonreciprocal \emph{and} lossless coupling, the $\mat\Omega$ matrix must take the form~\cite{WangAPB05}
\begin{equation}
  \label{eq:Omega-matrix}
  \mat\Omega = 
  \begin{bmatrix}
    \omega_0 & V\\ - V & \omega_0 
  \end{bmatrix},
\end{equation}
where $V\equiv \I \tilde V$ is purely imaginary. The real quantity~$\tilde V$ will henceforth be called the (MO) \emph{mode coupling strength}. Its form will be detailed in Section~\ref{ssec:cavity_design}. The eigenvalues of the above matrix, $\omega_\pm \equiv \omega_0 \pm \tilde V$, are the frequencies of the eigenmodes of the cavity in isolation (uncoupled to waveguides). The corresponding eigenvectors are obviously $(1,\pm\I)\transpose$, i.e.\ counterclockwise- and clockwise-rotating combinations of the even and odd cavity modes. The frequency splitting $\Delta \omega \equiv \abs{\omega_+ - \omega_-} = 2\abs{\tilde V}$ is proportional to the mode coupling strength. 

\subsection{Solution of the coupled-mode model}\label{ssec:solution_tcm}

We have now collected enough information to solve the coupled-mode equations \eqref{eq:suh-coupled-mode} and describe the device behaviour in terms of a minimal number of parameters. These are (1)~the frequency~$\omega_0$ of the degenerate cavity resonances, (2)~the MO mode coupling strength~$\tilde V$, (3)~the decay rate~$\gamma$ [related via Eq.~\eqref{eq:Gamma-matrix} to the cavity-port coupling strength], (4)~the phase difference~$\Delta+\pi$ between the scattering coefficients $r$ and~$t$, and finally (5)~the phase~$\tau$ of the direct pathway waveguide transmission coefficient $t$.

The formal solution of the coupled-mode equations~\eqref{eq:suh-coupled-mode} reads
\begin{equation}
\label{eq:formal_solution}
\vect s\tsub{out} = \bigl\{\mat C + \mat D [\I(\mat\Omega - \omega\mat I) + \mat\Gamma]^{-1}\mat D\transpose\bigr\}\vect s\tsub{in}.
\end{equation}
It is already apparent that device operation only depends quadratically on the coupling coefficient~$d$. Taking $W_1$ to be the input waveguide by setting $\vect s\tsub{in} = (1, 0, 0)\transpose$, and noting that $(D\transpose)_{*1}$, the first column of the matrix $D\transpose$, equals $d(1, 0)\transpose$, one obtains
\begin{equation}
\label{eq:formal-solution-2}
\vect s\tsub{out} = \mat C_{*1} + d^2 \bigl\{\mat{\tilde{D}} [\I(\mat\Omega - \omega\mat I) + \mat\Gamma]^{-1}\bigr\}_{*1}.
\end{equation}
The inverse of the matrix in brackets exists if and only if $\gamma \neq 0$ or $\omega \neq \omega_0 \pm \tilde V$, and is equal to
\begin{equation}
\label{eq:inversion}
\frac{1}{[\gamma - \I (\omega - \omega_0)]^2+\tilde V ^2}
\begin{bmatrix}
\gamma - \I (\omega - \omega_0) & \tilde V\\
-\tilde V & \gamma - \I (\omega - \omega_0)
\end{bmatrix}.
\end{equation}
We can now use Eqs.\ \eqref{eq:C-matrix}, \eqref{eq:r-and-t}, \eqref{eq:D-matrix-reduced}, \eqref{eq:intermediate-constraint-on-d}, \eqref{eq:t-r} and \eqref{eq:inversion} to express the quantities occurring in Eq.~\eqref{eq:formal-solution-2} in terms of the five parameters listed in the previous paragraph. This leads to 
\begin{widetext}
\begin{subequations}
  \begin{align}
    \label{eq:mph:model-scattering-coeffs-a}
    s\tsub{1,out} &= 
    \frac{\E^{\I\tau}}{\sqrt{1+8\cos^2\Delta}} 
    \biggl\{
    -\E^{\I\Delta} +
    \frac{4(3\cos\Delta + \I\sin\Delta)}{3}
    \frac{\gamma [\gamma - \I(\omega-\omega_0)]}
    {[\gamma - \I(\omega-\omega_0)]^2 + \tilde V^2}
    \biggr\},\\[\jot]
    \label{eq:mph:model-scattering-coeffs-b}
    s\tsub{2,out} &= 
    \frac{2\E^{\I\tau}}{\sqrt{1+8\cos^2\Delta}} 
    \biggl\{
    \cos\Delta -
    \frac{3\cos\Delta + \I\sin\Delta}{3}
    \frac{\gamma [\gamma +\tilde V \sqrt 3 -
      \I(\omega-\omega_0)]}
    {[\gamma - \I(\omega-\omega_0)]^2 + 
      \tilde V^2}
    \biggr\}, \\[\jot]
    \label{eq:mph:model-scattering-coeffs-c}
    s\tsub{3,out} &= 
    \frac{2\E^{\I\tau}}{\sqrt{1+8\cos^2\Delta}} 
    \biggl\{
    \cos\Delta -
    \frac{3\cos\Delta + \I\sin\Delta}{3}
    \frac{\gamma [\gamma -\tilde V \sqrt 3 -
      \I(\omega-\omega_0)]}
    {[\gamma - \I(\omega-\omega_0)]^2 + 
      \tilde V^2}
    \biggr\}.
  \end{align}
  \label{eq:mph:model-scattering-coeffs}
\end{subequations}
\end{widetext}

The transmission amplitudes $s\tsub{2,out}$ and $s\tsub{3,out}$ transform into each other when the sign of the MO coupling strength $\tilde V$ is inverted. This is consistent with the fact that $\tilde V$ is proportional to the magnetization of the MO material, as will be proven in Section \ref{ssec:cavity_design}, and that, as an axial vector, magnetization is inverted when mirrored in a plane parallel to it~\cite{Visnovskybook}. As a result the $W_3$-output of the ring circulator in Fig.~\ref{fig:geometry-general} must equal the (mirrored) $W_2$-output \emph{with opposite magnetization} (\emph{and thus~$\tilde V$}), and vice versa. It is easily seen that the reflectance $R \equiv \abs{s_{1,\mathrm{out}}}^2$ and transmittances $T_i \equiv \abs{s_{i,\mathrm{out}}}^2$ ($i = 2,3$) are independent of~$\tau$ and invariant under the mapping $(\Delta, \omega - \omega_0) \to (-\Delta, \omega_0 - \omega)$, which amounts to conjugating the expressions in curly brackets in Eqs.~\eqref{eq:mph:model-scattering-coeffs}.
Therefore it is sufficient to study the properties of $R$ and~$T_i$ in the interval $\Delta \in [0, \frac12\pi]$.

Apart from a global phase factor~$\tau$, the device's spectral behaviour can be normalized to the mode coupling strength $\tilde V$ and described entirely by three real parameters: (1) the reduced operation frequency $(\omega - \omega_0)/\tilde V$, (2) the reduced cavity-port decay rate $\gamma/\tilde V$, and (3) the phase lag $\Delta +\pi$ between the reflection $r$ and transmission~$t$ of the direct waveguide-to-waveguide coupling. The role of the latter is better understood by remarking that in Eq.~\eqref{eq:mph:model-scattering-coeffs} $\cos\Delta = \abs{t}/(2\abs{r})$. This direct coupling term, when present, adds an extra interfering contribution to both circulator transmittances, $T_2$ and $T_3$.

The spectral response of the cavity ports can be better understood by partial fraction decomposition of the second term between curly brackets in Eqs.~\eqref{eq:mph:model-scattering-coeffs}. It is straightforward to show that
\begin{subequations}
\label{eq:partialfracdecomp}
\begin{align}
\nonumber
s\tsub{2,out} &= 
    \E^{\I\tau} 
    \biggl\{\underbrace{
    \frac{2\cos\Delta}{\sqrt{1+8\cos^2\Delta}}}_{F(\Delta)} -
    \underbrace{\frac23\E^{\I \arg(3\cos\Delta+\I\sin\Delta)}}_{\phi(\Delta)}
    \\
    \label{eq:partialfracdecomp-a}
    &\quad \times\biggl[\underbrace{\frac{\gamma \E^{+\I\pi/3}}{\gamma - \I(\omega - \omega_+)}
      + \frac{\gamma \E^{-\I\pi/3}}{\gamma - \I(\omega - \omega_-)}}_{L_{\pi/3}(\omega-\omega_+) + L_{-\pi/3}(\omega-\omega_-)} 
    \biggr]
    \biggr\},\\[\jot]
    \nonumber
    s\tsub{3,out} &= 
    \E^{\I\tau} 
    \biggl\{
    \frac{2\cos\Delta}{\sqrt{1+8\cos^2\Delta}} -
    \frac23\E^{\I \arg(3\cos\Delta+\I\sin\Delta)}\\
    \label{eq:partialfracdecomp-b}
    &\quad\times
    \biggl[\frac{\gamma \E^{-\I\pi/3}}{\gamma - \I(\omega - \omega_+)}
    + \frac{\gamma \E^{+\I\pi/3}}{\gamma - \I(\omega-\omega_-)} 
    \biggr]
    \biggr\},
\end{align}
\end{subequations}
When the ports are perfectly decoupled ($\Delta = \pm\frac12\pi$), the circulator transmission coefficients are sums of standard Lorentzian lineshapes, $L_{\pm\pi/3}$, centred at the (split) resonant frequencies~$\omega_\pm$. In all other cases the direct term $F(\Delta)$ adds a continuum background coupling (with zero phase) and will thus in interaction with the Lorentzians add a sharp asymmetric Fano resonance to the circulator transmittance close to $\omega_\pm$. Fano resonances in integrated photonic structures have been predicted and observed in many configurations (see for instance Section~V in the review paper Ref.~\onlinecite{MiroshnichenkoRevModPhys2010}). The additional presence of the coupling term $\tilde V$ introduces an extra nonreciprocal asymmetry in the port transmittances. This results in the $\pm\frac13\pi$ phase shifts of the Lorentzians at the frequencies $\omega_\pm$ and in particular in the sign reversal of these phase shifts between the output ports. The impact of nonreciprocal coupling on Fano resonances in optical cavities is, up to our knowledge, studied here for the first time.

\begin{figure*}[!tb]
  \centering
  \includegraphics{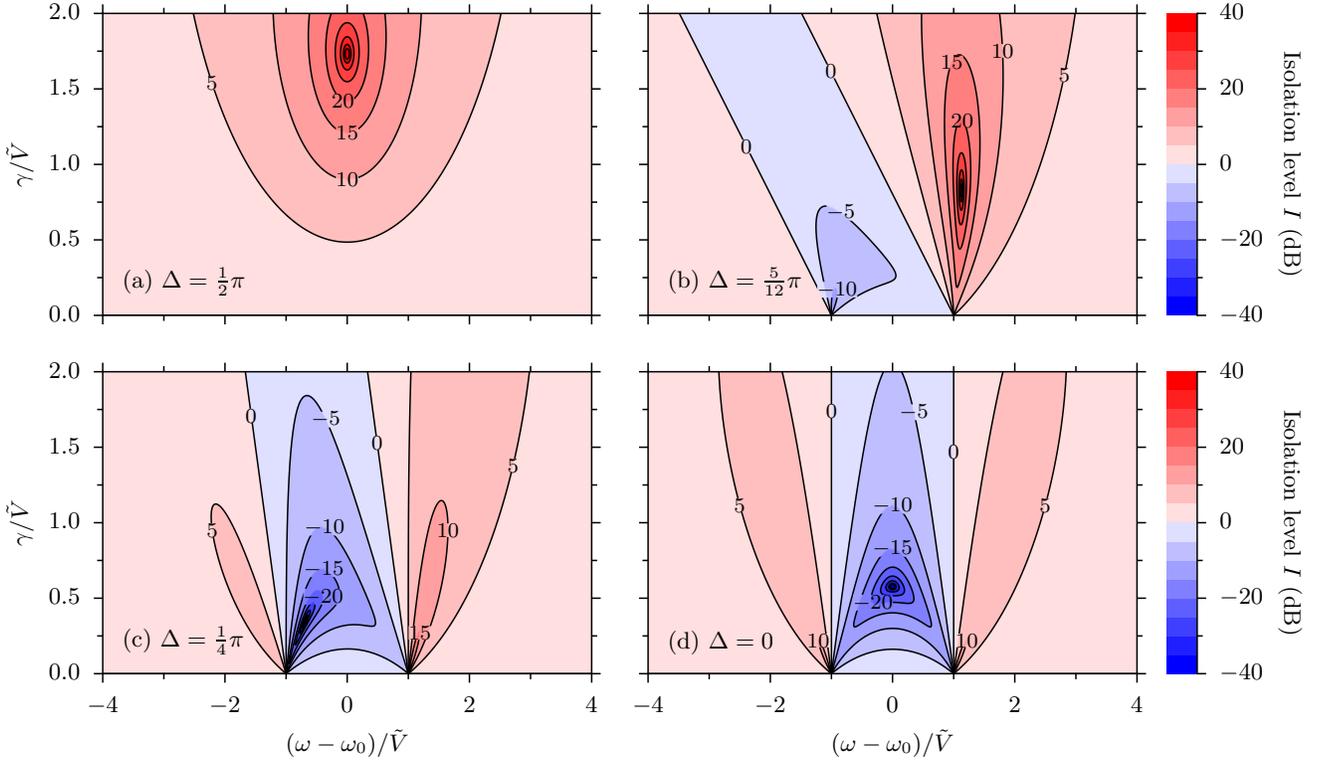}
  \caption{Isolation level~$I$ (in decibels) as a function of the reduced frequency shift $(\omega - \omega_0)/\tilde V$ and the reduced decay rate $\gamma/\tilde V$ for $\Delta = \frac12\pi$, $\frac{5}{12}\pi$, $\frac14\pi$ and $0$. The isolation contours are plotted every 5\,dB.}
  \label{fig:isolation-maps}
\end{figure*}

%%%%%%%%%%%%%%%%%%%%%%%%%%%%%%%%%%%%%%%%%%%%%%%%%%%%%%%%%%%%%%%%%%%%%%%%%%%%%%%%%%%%%%%%%%%%%%
\subsection{Discussion: Ideal lossless case}\label{ssec:discussion_tcm}
%%%%%%%%%%%%%%%%%%%%%%%%%%%%%%%%%%%%%%%%%%%%%%%%%%%%%%%%%%%%%%%%%%%%%%%%%%%%%%%%%%%%%%%%%%%%%%%

We shall now demonstrate the influence of these nonreciprocal Fano-type resonances on the \emph{isolation level} of the circulator, defined as
\begin{equation} 
I \equiv 10\log\frac{T_2}{T_3}.
\end{equation}
Figure~\ref{fig:isolation-maps} shows the dependence of~$I$ on the reduced frequency shift $(\omega - \omega_0)/{\tilde V}$ and the reduced decay rate~$\gamma/\tilde V$ for a few representative values of~$\Delta$. Several features can be distinguished in these plots. First of all, the spectral behaviour of the isolation level is strongly asymmetric around the resonant frequency $\omega_0$ of the cavity, except for the limiting cases of zero ($\Delta=\frac12\pi$) and maximum ($\Delta=0$) direct coupling. This is obviously a signature of Fano interferences. Secondly, regardless of the level of direct coupling there always seems to exist a frequency for which infinite isolation (i.e. $T_2 = 0$ or $ T_3 = 0$) can be obtained by properly adjusting the cavity-port coupling strength~$\gamma$. In fact, it can be proven mathematically using Eqs.\ \eqref{eq:mph:model-scattering-coeffs-b} and~\eqref{eq:mph:model-scattering-coeffs-c} that perfect isolation can indeed occur at all values of $\Delta$ except $\pm\frac13\pi$. Thirdly, the sign of the maximum (infinite) isolation level varies as a function of $\Delta$. This implies that by ``tuning'' the level of direct waveguide-to-waveguide coupling the sense of the circulation can be reversed! Finally, the circulation also exhibits a spectral sign change as soon as direct coupling appears. As a result, the sense of circulation (clockwise or counterclockwise) changes outside a certain band of frequencies around the optimum frequency. This might have interesting applications for integrated optical add-drop functions.

In practical applications one is also concerned about maximizing the bandwidth $B(I\tsub{min})$, defined as the length of the frequency interval in which the magnitude of the isolation level exceeds a predetermined threshold $I\tsub{min}$. Figure~\ref{fig:isolation-maps} shows that this bandwidth is particularly large for vanishing ($\Delta = \frac12\pi$) and maximum ($\Delta = 0$) direct waveguide-to-waveguide coupling. In these cases one proves easily using Eqs.~\eqref{eq:mph:model-scattering-coeffs-b} and \eqref{eq:mph:model-scattering-coeffs-c} that $T_2$ and $T_3$ are even functions of~$(\omega - \omega_0)$ and, thus, have an extremum at $\omega_0$. Perfect isolation will therefore occur at $\omega = \omega_0$, provided that the cavity-port coupling level~$\gamma$ takes a specific value $\gamma_\infty$, equal to $\abs{\tilde V}\sqrt 3$ for $\Delta = \frac12\pi$ or to $\abs{\tilde V}/\sqrt 3$ for $\Delta = 0$. Interestingly, however, the bandwidth $B(I\tsub{min})$ for a given value of $I\tsub{min}$ can sometimes be increased by tuning $\gamma$ away from $\gamma_\infty$. For instance, Fig.\ \ref{fig:isolation-maps}(d) shows that when $\Delta = 0$, the 20-dB bandwidth is largest for $\gamma$ slightly smaller than $\gamma_\infty$. In fact, Eqs.\ \eqref{eq:mph:model-scattering-coeffs-b} and~\eqref{eq:mph:model-scattering-coeffs-c} can be used to show that the maximum bandwidths $B(I\tsub{min})$ obtainable in circulators with $\Delta = 0$ and $\Delta = \frac12\pi$ are \emph{identical} for all $I\tsub{min} \geq 10\log(31 + 8 \sqrt{15}) \approx 18$\,dB. Nonetheless, the structure with zero direct coupling ($\Delta = \frac12 \pi$) has the advantage of lesser sensitivity to perturbations of the cavity-port coupling level off its optimum value. In addition, for this structure the value of $\gamma$ maximizing the bandwidth for any reasonably high value of $I\tsub{min}$ is very close to $\gamma_\infty$, the value at which perfect circulation becomes possible. This facilitates the device design.

\begin{figure*}[!htb]
  \centering
  \includegraphics{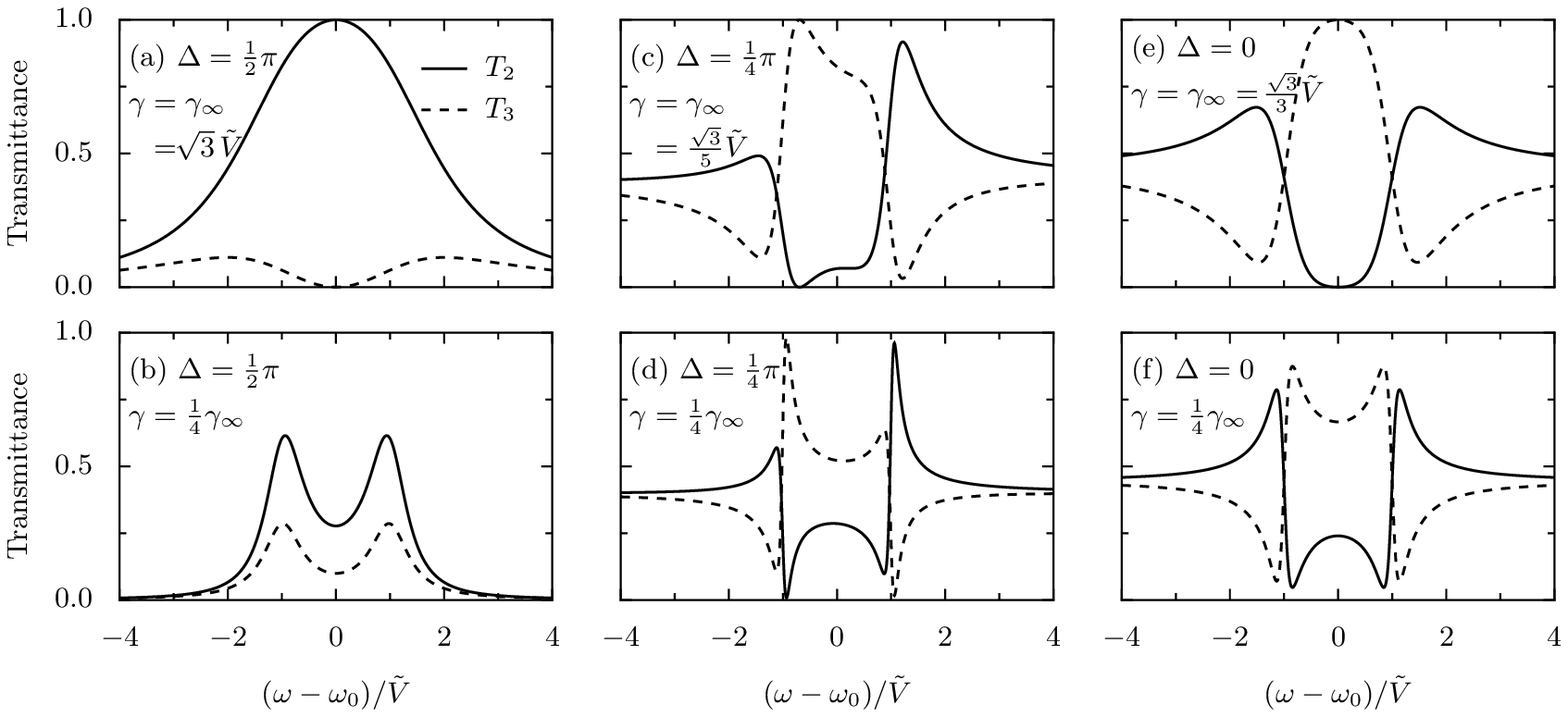}
  \caption{Frequency dependence of the port transmittances following Eqs.~\eqref{eq:mph:model-scattering-coeffs-b} and \eqref{eq:mph:model-scattering-coeffs-c} for \subnumber{(a)}--\subnumber{(b)} complete absence of direct pathway coupling ($\Delta = \frac12\pi$); \subnumber{(c)}--\subnumber{(d)} an intermediate level of direct coupling ($\Delta = \frac14\pi$); and \subnumber{(e)}--\subnumber{(f)} maximum direct pathway coupling ($\Delta=0$). For each considered $\Delta$, the lower plot corresponds to optimal $\gamma = \gamma_\infty$ which gives either $T_2$ or $T_3 = 0$ at a certain frequency. The upper one shows the behaviour when $\gamma$ is suboptimal, $\gamma=\frac14\gamma_\infty$.}
  \label{fig:Fano}
\end{figure*}

In Fig.~\ref{fig:Fano} we have plotted the frequency dependence of the port transmittances separately for three scenarios: complete absence of direct coupling ($\Delta = \frac12\pi$), maximum direct coupling ($\Delta = 0$), and an intermediate level of direct coupling ($\Delta=\frac14\pi$). For each of the three scenarios, the port transmittances are plotted for two levels of the cavity-port coupling: $\gamma = \gamma_{\infty}$ (top) and $\gamma = \frac14\gamma_{\infty}$ (bottom). The features mentioned above appear clearly. For instance, the sign change of the isolation level --- $T_2 > T_3 \rightarrow T_3 > T_2$ --- when going from perfectly decoupled to maximally coupled circulator ports is clearly visible. Likewise, as soon as direct coupling appears ($\Delta \neq \pm\frac12\pi$), one can observe how Fano resonances in the port transmittances are responsible for both the asymmetry and the spectral sign changes of the isolation levels.

\begin{figure*}%
\centering\includegraphics{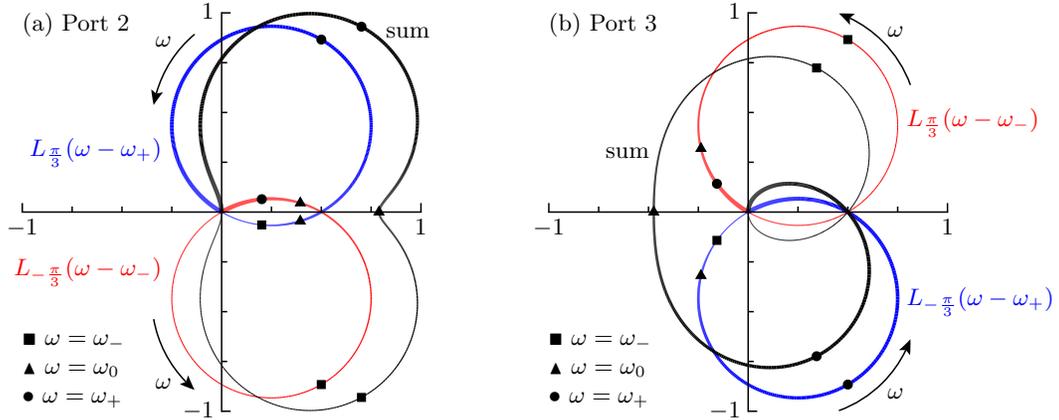}
\caption{Traces of the two complex Lorentzians occurring in \subnumber{(a)}~Eq.\ \eqref{eq:partialfracdecomp-a} for $s_{2,\mathrm{out}}$ and \subnumber{(b)}~Eq.\ \eqref{eq:partialfracdecomp-b} for $s_{3,\mathrm{out}}$ (with $\gamma = \frac14 \gamma_\infty = \frac{\sqrt3}4 \tilde V$) treated as complex functions of the frequency $\omega$. As $\omega$ goes from $-\infty$ to $\infty$, $L_{\pi/3}(\omega-\omega_+)$ and $L_{-\pi/3}(\omega-\omega_-)$ follow the blue and red circular paths, respectively, in the direction of increasing line thickness, starting and ending at the origin. The black curve traces the sum of the two Lorentzians. The rate of change of the line thickness, as well as the locations of the markers corresponding to $\omega_-$, $\omega_0$ and $\omega_+$, indicate the speed with which the curves are traversed.}%
\label{fig:Fanocomplex-lorentzians}%
\end{figure*}

\begin{figure*}%
\centering\includegraphics{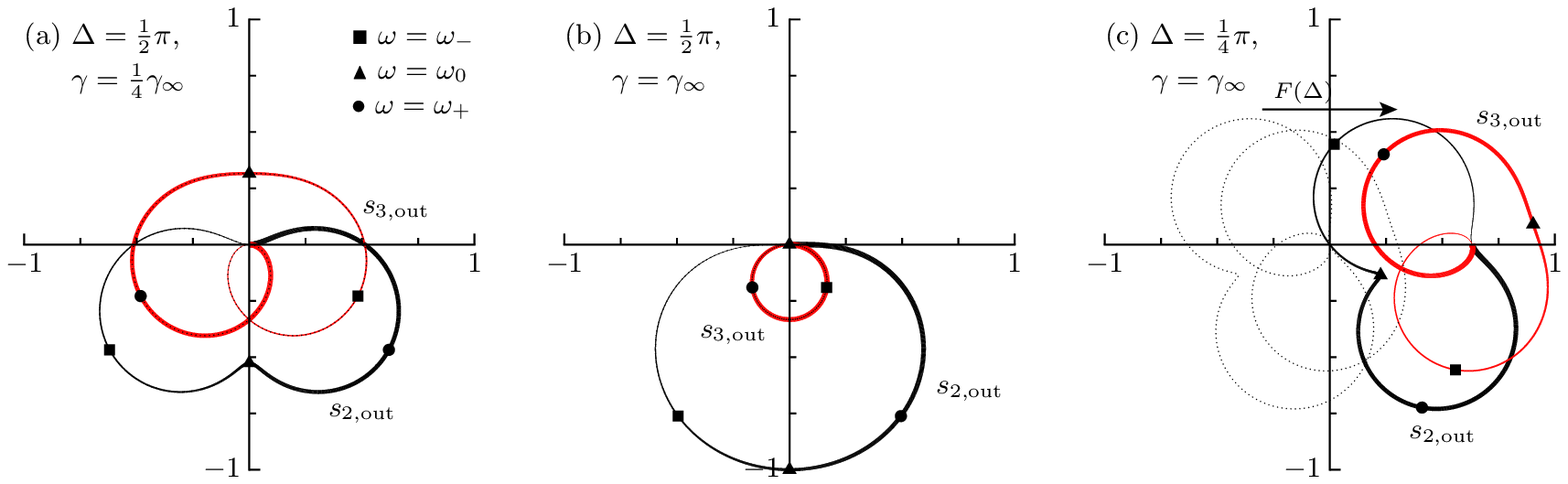}
\caption{\subnumber{(a)}~Traces of $s\tsub{2,out}$ and $s\tsub{3,out}$, treated as complex functions of the frequency $\omega$, of a circulator with no direct-pathway port coupling ($\Delta = \frac12\pi$), a suboptimal cavity-port coupling strength $\gamma = \frac14\gamma_\infty = \frac{\sqrt{3}}4\tilde V$ and phase $\tau = 0$. Increasing line thickness indicates as before the evolution of $\omega$ from $-\infty$ to $\infty$. Note that because of the absence of direct coupling, the two paths are simply the black curves from Figs.\ \ref{fig:Fanocomplex-lorentzians}(a) and (b) scaled by the complex factor $\phi(\Delta=\frac12\pi) = \frac23\E^{-\I\pi/2}$. Note also that Fig.~\ref{fig:Fano}(b) shows the squared modulus of these parametric complex curves. \subnumber{(b)}~Same for the optimum cavity-port coupling strength $\gamma = \gamma_\infty$. The black curve is traversed twice. The squared moduli of these curves correspond to Fig.~\ref{fig:Fano}(a). \subnumber{(c)}~Same for $\Delta = \frac14\pi$ and $\gamma = \gamma_\infty = \frac{\sqrt3}5\tilde V$. The dotted curves show $s_{i,\mathrm{out}}$ after substracting the direct-coupling term $F(0)$. Perfect circulation occurs now at a frequency different from $\omega_0$. This case corresponds to Fig.~\ref{fig:Fano}(c).}%
\label{fig:Fanocomplex-complete}%
\end{figure*}

The appearance of the above features can be better understood by considering the form of Eqs.~\eqref{eq:partialfracdecomp} for the complex transmission coefficients. Neglecting the global phase factor $\E^{\I\tau}$, they contain two contributions: (1)~a constant real term $F(\Delta)$ that varies from~0 to~$\frac23$ as $\Delta$~goes from~$\frac12\pi$ to~$0$ and (2)~the sum of two oppositely $\frac13\pi$-phase-shifted Lorentzians. The latter complex sum is multiplied by the factor $-\phi(\Delta)$ of constant magnitude, $\frac23$, and phase that varies from $\frac12\pi$ to $0$. As shown in Fig.~\ref{fig:Fanocomplex-lorentzians}, in the complex plane the Lorentzian parametric curves describe circles with radius~$\frac12$, starting and ending at the origin and attaining their ``apex'' at $\E^{\pm\I\pi/3}$ when $\omega = \omega_\pm$, respectively. For the other port, the roles of the circles are interchanged. When these two parametric circular curves are added, one obtains the loci shown with the black curves in Fig.~\ref{fig:Fanocomplex-lorentzians}. In the absence of direct-pathway coupling, the Fano term (a) is zero and these curves, multiplied by the complex factor $-\phi(\frac12\pi) = -\frac23\I$, effectively describe the port transmission coefficients. For a suboptimal cavity-port coupling strength~$\gamma$, at the central frequency $\omega = \omega_0$  the values of the two Lorentzians contributing to $s\tsub{3,out}$ [marked with triangles in Fig.\ \ref{fig:Fanocomplex-lorentzians}(a)] do not exactly cancel each other. As a result, as shown in Fig.~\ref{fig:Fanocomplex-complete}(a), complete isolation is not achieved. By adjusting the value of $\gamma$ to $\gamma_\infty = \tilde V\sqrt3$, the speed with which the two circles are traversed can be tuned so that at $\omega=\omega_0$ the two Lorentzians take opposite, purely imaginary values, thus making $s\tsub{3,out}$ vanish. This situation is illustrated by Fig.~\ref{fig:Fanocomplex-complete}(b). Fig.~\ref{fig:Fanocomplex-complete}(c) shows a similar construction for an intermediate level of direct coupling ($\Delta=\frac14\pi$). This description makes the behaviour of the port transmittances for $\Delta \neq \pm\frac12\pi$ easier to understand. The complex factor $\phi(\Delta)$ first ``rotates'' the complex paths $L_{\pi/3}(\omega-\omega_\pm) + L_{-\pi/3}(\omega-\omega_\mp)$ to the positions indicated by the dotted curves in Fig.~\ref{fig:Fanocomplex-complete}(c). The Fano interference term $F(\Delta)$ subsequently shifts the paths along the real axis. The latter shift explains why the ratio $\log(\abs{s_2}/\abs{s_3})$ can change sign as a function of frequency: the paths do not start or end anymore in the origin of the complex plane. On the other hand, the asymmetric phase shift induced by $\phi(\Delta)$ is responsible for the sharp asymmetries of the port transmittances. Only for maximal direct coupling does the symmetry reappear, since $\arg(-\phi(0))=\pi$.

In summary, an abstract coupled-mode-theoretical description of a 3-port circulator including possible Fano resonances shows that the circulator ports do not necessarily have to be perfectly decoupled (as in Refs.\ \onlinecite{SmiOL10,WangOL05}) in order to obtain reasonable behaviour. In fact, somewhat paradoxically, devices with maximum direct coupling also provide a large bandwidth, exhibiting, in addition, an interesting switching effect. This is a first indication that a technologically less demanding cavity layout not embedded in a PhC background --- and hence without a guarantee of perfect port decoupling --- is worth considering.

%%%%%%%%%%%%%%%%%%%%%%%%%%%%%%%%%%%%%%%%%%%%%%%%%%%%%%%%%%%%%%%%%%%%%%%%%%%%%%%%%%%%%%%%%%%%%%%%%%%%%%%%%%%%%%%%%%%%%%%%%%%%%%%%%%%%%%%%%%%%%%

\subsection{Inclusion of radiation loss}\label{ssec:tcm_radiation}

Another, probably more detrimental, effect that can play an important part in the operation of circulators is the radiative energy loss. It can be fairly easily handled by the coupled-mode theory. We will limit our discussion to the case without direct coupling ($\Delta = \frac12\pi$), which in the previous subsection was found to yield maximum bandwidth and to be most tolerant to perturbations of the cavity-port coupling strength. We will also assume the energy to be lost solely owing to radiation of the cavity mode into free space, neglecting the energy radiated when the incident waveguide mode is reflected from the end of the input waveguide. This allows us to assume that the form of the coupling matrices $\mat C$ and $\mat D$ is unaffected. Note that Eq.~\eqref{eq:suh-constraint-1} remains valid if by $\mat \Gamma$ in this equation one implies only the decay rate to the waveguide ports.

Following Ref.\ \onlinecite{Joannopoulosbook2nded}, pp.\ 208--212, we incorporate the cavity radiation loss in the coupled-wave equations~\eqref{eq:suh-coupled-mode} by replacing~$\mat\Gamma$ by $\mat\Gamma + \mat\Gamma\tsub r$, where $\mat\Gamma\tsub r \equiv \gamma\tsub r \bigl[\begin{smallmatrix}1&0\\0&1\end{smallmatrix}\bigr]$ and $\gamma\tsub r$~is the decay constant related to radiation loss. Proceeding analogously as before, we solve the coupled-wave equations for the amplitudes of the outgoing modes in the three waveguides:
\begin{subequations}
  \label{eq:loss:scattering-coeffs}
  \begin{align}
    \label{eq:loss:scattering-coeffs-a}
    s\tsub{1,out} &= 
    -1 + 
    \frac43
    \frac{\gamma [\gamma +\gamma\tsub r - \I(\omega-\omega_0)]}
    {[\gamma +\gamma\tsub r - \I(\omega-\omega_0)]^2 +
      \tilde V^2},\\[\jot]
    \label{eq:loss:scattering-coeffs-b}
    s\tsub{2,out} &= 
    -\frac23
    \frac{\gamma [\gamma + \gamma\tsub r +\tilde V \sqrt 3 -
      \I(\omega-\omega_0)]}
    {[\gamma +\gamma\tsub r - \I(\omega-\omega_0)]^2 + \tilde V^2},
    \\[\jot]
    \label{eq:loss:scattering-coeffs-c}
    s\tsub{3,out} &= 
    -\frac23
    \frac{\gamma [\gamma + \gamma\tsub r -\tilde V \sqrt 3 -
      \I(\omega-\omega_0)]}
    {[\gamma +\gamma\tsub r - \I(\omega-\omega_0)]^2 + \tilde V^2}.
  \end{align}
\end{subequations}
From eqs.\ \eqref{eq:loss:scattering-coeffs-b} and \eqref{eq:loss:scattering-coeffs-c} it can be immediately seen that despite the presence of loss, infinite isolation level~$I$ can still be obtained at frequency $\omega = \omega_0$ provided that the coupling coefficient $\gamma$ is taken as
\begin{equation}
  \label{eq:loss:gamma}
  \gamma = \abs{\tilde V} \sqrt 3 - \gamma\tsub r.
\end{equation}
This underlines that in the presence of radiation loss in the cavity the critical coupling of the waveguide ports to the cavity to obtain perfect isolation must become weaker. If the radiative decay rate $\gamma\tsub r$ becomes too high, the above condition cannot be fulfilled for any positive value of $\gamma$ and hence perfect circulation becomes impossible. Similar conclusions can be reached if the waveguide ports are not perfectly decoupled ($\Delta \neq \frac12\pi$). Interestingly, for the case of $\Delta = \frac12\pi$, when $\gamma$ is chosen along\ \eqref{eq:loss:gamma}, the isolation level $I(\omega)$ [and the bandwidth $B(I_\mathrm{min})$] becomes independent from $\gamma\tsub r$:
\begin{subequations}
  \label{eq:loss:I(omega)}
\begin{align}
  I(\omega) &= 10\log\biggl[1 + \frac{12\tilde V^2}{(\omega - \omega_0)^2}\biggr]\\
\intertext{and}
  B(I\tsub{min}) &= \frac{4\sqrt{3}\abs{\tilde V}}{\sqrt{10^{I\tsub{min}/10}-1}}.
\end{align}
\end{subequations}
However, the power transmitted to the `hot' waveguide, $T_3$, obviously decreases when the loss coefficient~$\gamma\tsub r$ increases. With $\gamma$ given by Eq.~\eqref{eq:loss:gamma}, the expression for the maximum value of~$T_3$ takes the form
\begin{equation}
  \label{eq:loss:max-transmission}
  T_3(\omega = \omega_0) = \frac{(\sqrt3 - \gamma\tsub r / \abs{\tilde V})^2}3.
\end{equation}

This formula is important because it allows to determine the minimum quality factor $Q\tsub r \equiv \omega_0 / (2\gamma\tsub r)$ of a cavity with a given relative frequency splitting $\Delta\omega/\omega_0$ that can be used to build a circulator with a prescribed minimum level of transmitted power, $T\tsub{3,min}$. (Note that $Q\tsub r$ describes solely the decay caused by radiation to free space, rather than that due to interaction with waveguide modes in the circulator ports.) Indeed, expressing $\gamma\tsub r$ in Eq.~\eqref{eq:loss:max-transmission} in terms of $Q\tsub r$ and using the relation $\Delta\omega = 2\abs{\tilde V}$, we obtain that 
$T_3(\omega = \omega_0) \geq T\tsub{min}$ if and only if
\begin{equation}
  \label{eq:loss:min-Q-l}
%   Q\tsub r \geq \biggl[\frac{2 \sqrt 3 \tilde V}{\omega_0}
%   (1 -\sqrt{T\tsub{min}})\biggr]^{-1}.
  Q\tsub r \geq \biggl[\frac{\Delta\omega}{\omega_0}
  (1 -\sqrt{T\tsub{min}})\sqrt 3\biggr]^{-1}.
\end{equation}
Thus, a circulator built with a cavity having $\Delta\omega / \omega_0 = 0.001$ (a typical value) can provide 50-percent peak transmission if $Q\tsub r \geq 1970$. A quality factor $Q\tsub r \geq 11{,}250$ is needed for 90-percent efficiency, and $Q\tsub r \geq 115{,}000$ for 99-percent.

\declarecirculator{circ:best-rectangular} % C22
\declarecirculator{circ:best-conical} % C27

%%%%%%%%%%%%%%%%%%%%%%%%%%%%%%%%%%%%%%%%%%%%%%%%%%%%%%%%%%%%%%%%%%%%%%%%%%%%%%%%%%%%%%%%%%%%
%
\section{Rib-waveguide-based circulators}\label{sec:circulator_design}
%
%%%%%%%%%%%%%%%%%%%%%%%%%%%%%%%%%%%%%%%%%%%%%%%%%%%%%%%%%%%%%%%%%%%%%%%%%%%%%%%%%%%%%%%%%%%%%%%

%%%%%%%%%%%%%%%%%%%%%%%%%%%%%%%%%%%%%%%%%%%%%%%%%%%%%%%%%%%%%%%%%%%%%%%%%%%%%%%%%%%%%%%%%%%%%%%%
%
\subsection{Design of axisymmetric ring cavities}\label{ssec:cavity_design}
%
%%%%%%%%%%%%%%%%%%%%%%%%%%%%%%%%%%%%%%%%%%%%%%%%%%%%%%%%%%%%%%%%%%%%%%%%%%%%%%%%%%%%%%%%%%%%%%%%%%
As is clear by now, the performance of a cavity-based circulator depends heavily on the strength~$\tilde V$ of the MO coupling of the even and odd cavity modes. As can be seen from Eqs.\  \eqref{eq:loss:gamma} and \eqref{eq:loss:I(omega)}, an increase of $\abs{\tilde V}$ augments the bandwidth, necessitates less critical coupling and leads to higher tolerance towards radiation loss. 
Wang and Fan~\cite{WangAPB05} derived a perturbational formula for the mode coupling strength~$\tilde V$ of a cavity containing a MO material polarized in the $z$~direction. Such a material is characterized by the permittivity tensor
\begin{equation}
  \label{eq:mo-epsilon-z-pol}
  \mat\epsilon = 
  \begin{bmatrix}
    \epsilon & \I g & 0 \\
    -\I g & \epsilon & 0 \\
    0 & 0 & \epsilon
  \end{bmatrix},
\end{equation}
where $g$, proportional to the material's magnetization, is the gyroelectric index. From now on we shall assume~$g$ to be real. Together with the real-valuedness of~$\epsilon$ this is a necessary condition for a Hermitian $\mat\epsilon$ and thus a lossless MO material. Wang and Fan~\cite{WangAPB05} obtained~\footnote{The difference in sign between Eq.\ (6) from Ref.\ \onlinecite{WangAPB05} and Eq.~\eqref{eq:wang-V} here is caused by the difference in the  convention chosen for harmonic time-dependence ($\E^{-\I\omega t}$ here vs.\ $\E^{\I\omega t}$ in Ref.\ \onlinecite{WangAPB05}).}
\begin{equation}
  \label{eq:wang-V}
  V \equiv \I\tilde V = -\frac\I2 
  \frac{\omega_{0} \int g(\vect r) \,\vers z \cdot 
    [\vect E\tsub e^*(\vect r) \times \vect E\tsub o(\vect r)] \,\diff\vect r}
  {\sqrt{
      \int \epsilon(\vect r) \abs{\vect E\tsub o(\vect r)}^2 \, \diff \vect r
      \int \epsilon(\vect r) \abs{\vect E\tsub e(\vect r)}^2 \, \diff \vect r}},
\end{equation}
where the integrals run over the whole cavity volume, $\vect E\tsub e$ and~$\vect E\tsub o$ are the electric fields of the even and odd modes of the non-magnetized cavity, and $\omega_0$ is their frequency. Thus, $V$~is proportional to the cross product of $\vect E\tsub e^*$ and~$\vect E\tsub o$ weighted by the off-diagonal component of the permittivity tensor, $\I g$.

 It is convenient to introduce the dimensionless \emph{reduced} MO coupling strength $\tilde v$ defined as $\abs{\tilde V}/(\omega_0 \abs{g}\tsub{max})$, where $\abs{g}\tsub{max}$ denotes the maximum magnitude of~$g$ over the cavity volume. The parameter~$\tilde v$ depends then solely on the geometry of the cavity and not on the strength of the MO effect in the chosen material. It can therefore be used to compare the merits of cavities with different geometries.

By exploiting the analytical properties of the eigenmodes of the Helmholtz equation in axisymmetric geometries, we will now derive a general design strategy for resonant 2D cavities with large frequency splitting in a \emph{uniform} static external magnetic field. In contrast to general PC cavities, circularly symmetric cavities can be handled analytically, which makes it possible to get a better insight into their properties.

\begin{figure}[!tb]
  \centering
  \includegraphics{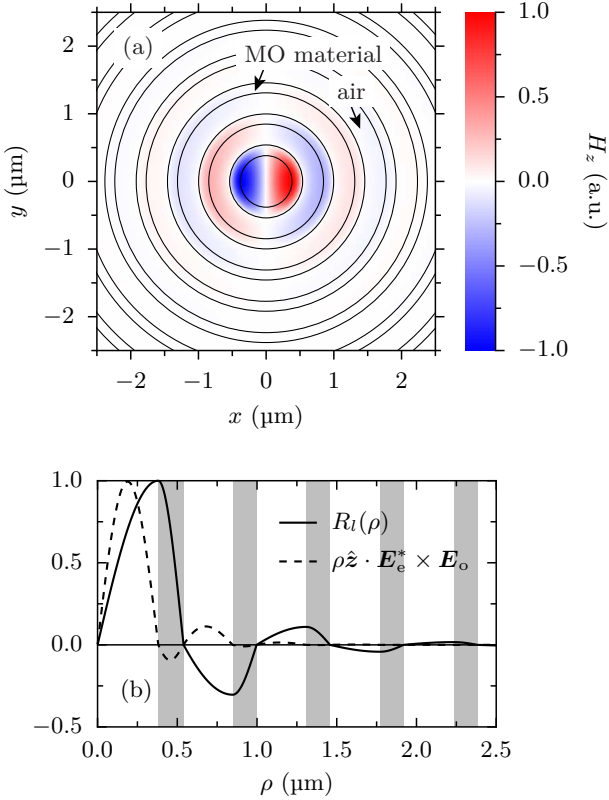}

  \caption{\subnumber{(a)}~Magnetic field $H_z$ of the even eigenmode of azimuthal order $l = 1$ and wavelength $\lambda = 1.300$\,\textmu m supported by the cavity composed of an alternating sequence of rings made of air and a MO material with refractive index 2.25. \subnumber{(b)}~Radial dependence of the magnetic field $H_z$ of the cavity eigenmodes and of the cross product of their electric fields. Both functions are normalized to their maximum values. The areas filled with the MO material are shaded.}
  \label{fig:circular-cavity}
\end{figure}

We consider, then, a system of concentric rings [see Fig.~\ref{fig:circular-cavity}(a)], described by a piecewise-constant relative permittivity $\epsilon(\rho )$ independent from the azimuthal coordinate~$\phi$ and the vertical coordinate~$z$. The relative permeability~$\mu$ is taken to be 1 everywhere. In the $p$-polarization case [waves propagating in the $(\rho , \phi)$ plane with the magnetic field parallel to the $z$~axis] Maxwell's equations 
\begin{subequations}
  \label{eq:maxwell}
  \begin{align}
    \frac{1}{\rho } \pderiv{H_z}{\phi} &= -\I\omega\epsilon\epsilon_0 E_\rho ,\\
    \pderiv{H_z}{\rho } &= \I\omega\epsilon\epsilon_0 E_\phi,\\
    \pderiv{E_\phi}{\rho } + \frac{1}{\rho }E_\phi - \frac{1}{\rho } \pderiv{E_\rho }{\phi} &=
    \I\omega\mu_0 H_z
  \end{align}
\end{subequations}
reduce to the Helmholtz equation for the $z$ component of the magnetic field, $H_z$:
\begin{equation}
  \label{eq:helmholtz}
  \epsilon \pderiv{}{\rho } \biggl(\frac{1}{\epsilon} \pderiv{H_z}{\rho }\biggr) +
  \frac{1}{\rho } \pderiv{H_z}{\rho } + \frac{1}{\rho ^2} \pderiv[2]{H_z}{\phi} +
  \epsilon \frac{\omega^2}{c^2} H_z = 0.
\end{equation}
Here $\omega$ denotes the frequency, $\epsilon_0$ and $\mu_0$ the permittivity and permeability of free space,  and $c = 1/\sqrt{\epsilon_0\mu_0}$ the speed of light in vacuum. By the usual method of separation of variables one can show that the eigenmodes of the system occur in degenerate pairs of ``even'' and ``odd'' modes with magnetic field of the general form
\begin{equation}
  \label{eq:H_z-even-and-odd}
  H_{z\mathrm{e}}(\rho , \phi) = R_l(\rho ) \cos(l\phi)
  \quad\text{and}\quad
  H_{z\mathrm{o}}(\rho , \phi) = R_l(\rho ) \sin(l\phi),
\end{equation}
respectively, where the azimuthal order~$l$ is an integer and the radial dependence $R_l(\rho)$ is a solution of
\begin{equation}
\label{eq:design:BesselR}
\epsilon(\rho) \pderiv{}{\rho } \biggl[\frac{1}{\epsilon(\rho)} \pderiv{R_l}{\rho }\biggr] +
  \frac{1}{\rho} \pderiv{R_l}{\rho} + \epsilon(\rho)k_0^2\biggl[1-\frac{l^2}{\epsilon(\rho)k_0^2\rho^2}\biggr]R_l(\rho) = 0,
\end{equation} 
which reduces to the Bessel equation within each uniform ring. In the above equation $k_0 \equiv \omega/c$ is the free-space wave number. Within each ring, $R_l(\rho)$ can then be expressed as a superposition of the Bessel functions of the first and second kind:
\begin{equation}
  \label{eq:R_l}
  R_l(\rho ) = a_m J_l(n_m k_0 \rho ) + b_m Y_l(n_m k_0 \rho ),
\end{equation}
where $m$ is the ring's number, $n_m (\equiv \sqrt{\epsilon_m})$ its refractive index, and $a_m$ and $b_m$ constant coefficients.

From Eq.~\eqref{eq:H_z-even-and-odd} and Maxwell's equations~\eqref{eq:maxwell} we can derive the general expressions for the cross product of the electric fields corresponding to a pair of degenerate modes with frequency $\omega_0$ and azimuthal order~$l$,
\begin{equation}
  \label{eq:circular-cross-product}
  \vers z \cdot (\vect E\tsub e^* \times \vect E\tsub o) =
  \frac{l R_l(\rho ) R_l'(\rho )}{\omega_0^2 \rho\, \epsilon_0^2\,[\epsilon(\rho )]^2},
\end{equation}
and for their squared norm,
\begin{equation}
  \label{eq:circular-norm}
  \begin{split}
  \int \epsilon(\vect r) \abs{\vect E\tsub e(\vect r)}^2 \, \diff \vect r &=
  \int \epsilon(\vect r) \abs{\vect E\tsub o(\vect r)}^2 \, \diff \vect r \\
  &=
  \pi \int_0^\infty \frac{l^2 [R_l(\rho )]^2 + \rho ^2 [R_l'(\rho )]^2}{\omega_0^2\rho \,\epsilon_0^2\, \epsilon(\rho )} \,\diff \rho ,
  \end{split}
\end{equation}
with $R_l'(\rho )$ denoting the derivative of $R_l(\rho )$. Substituting Eqs.\ (\ref{eq:circular-cross-product}) and (\ref{eq:circular-norm}) to the general formula for the coupling strength~\eqref{eq:wang-V}, we obtain
\begin{equation}
  \label{eq:circular-V}
  V = 
  -\frac{\displaystyle \I\omega_0 l\int_0^\infty \frac{g(\rho ) R_l(\rho ) R_l'(\rho )}{\epsilon(\rho )^2} \,\diff \rho }
  {\displaystyle \int_0^\infty \frac{l^2 [R_l(\rho )]^2 + \rho ^2 [R_l'(\rho )]^2}{\rho \, \epsilon(\rho )} \,\diff \rho }.
\end{equation}
Clearly, the sign of the integrand in the numerator of the above expression depends on the sign of the product $g(\rho )R_l(\rho )R_l'(\rho )$. In general, this sign will oscillate due to the oscillatory behaviour of the function $R_l(\rho )$ itself. Therefore, to avoid unnecessary cancellations in the integral in question, the ring boundaries need to be placed so that the product $g (\rho)R_l(\rho )R_l'(\rho )$ be always nonnegative (or nonpositive). \emph{Short of introducing inversely-magnetized domains, this can only be achieved by ensuring that the boundaries between the MO and non-MO material coincide with the zeros of the product $R_l(\rho )R_l'(\rho )$, i.e., the zeros and extrema of $R_l(\rho )$.} With this design principle in mind, we offer the following method for the determination of the radii of the alternating MO and non-MO rings making up a cavity supporting a pair of degenerate modes with azimuthal order $l \geq 1$ at a particular frequency $\omega_0 = k_0 c$.

As a first step, we determine the radius~$\rho _0$ of the central rod (``zeroth ring''), assumed to have refractive index~$n_0$. In this rod, the radial field dependence is expressed solely by means of the Bessel function of the first kind $J_l(n_0 k_0 \rho )$, since $Y_l(n_0 k_0 \rho )$ has a singularity at the origin. Hence, $b_0 = 0$ and, since the global mode amplitude is arbitrary, $a_0$~can be set to~$1$. It is then evident that the product $R_l(\rho ) R_l'(\rho )$ in the central rod will not change sign until the first extremum of the $J_l(n_0 k_0 \rho )$ function. Therefore we set~$\rho _0$ to the value of~$\rho $ at which this extremum occurs.

The amplitudes $a_1$ and $b_1$ of the Bessel functions in the first ring, having refractive index $n_1$, can now be determined from the condition of continuity of $H_z$ and $E_\phi$. The outer radius of the first ring, $\rho _1$, should then be chosen so as to coincide with the first zero of the function $a_1 J_l(n_1 k_0 \rho ) + b_1 Y_l(n_1 k_0 \rho )$ located in that ring: this is where the product $R_l(\rho ) R_l'(\rho )$ will again change sign. By repeating this process, we obtain the radii of the subsequent rings. The outer radii of the even rings (with refractive index $n_0$) coincide with the extrema of the function $R_l(\rho )$, and those of the odd rings (with refractive index $n_1$), with its zeros. Thus, the cavity designed in this way resembles an annular Bragg grating~\cite{ScheuerJOSAB07}, with each layer a quarter-wavelength thick (in the sense of the Bessel-function quasi-periodicity).

We need now to make sure that the field of the mode constructed in this way is localized, i.e., that $R_l(\rho)$ decays as $\rho$ tends to infinity. Let us estimate the ratio of $R_l(\rho)$ at its two successive extrema, located at $\rho_{2n}$ and $\rho_{2n+2}$, where $n \gg 1$. The Bessel functions of large arguments can be approximated by \cite[Eqs.\ (9.2.1) and (9.2.2)]{Abramowitz1965}
\begin{equation}
  \label{eq:asymptotic-bessel}
    \begin{Bmatrix} J_l(x)\\[\jot] Y_l(x) \end{Bmatrix}
    \approx \sqrt{\frac{2}{\pi x}} 
    \begin{Bmatrix} \cos \\[\jot] \sin \end{Bmatrix}
    \biggl(x-\frac\pi 2 l -\frac \pi 4\biggr).
\end{equation}

By construction, the field in the $(2n+1)$th ring, located between $\rho_{2n}$ and $\rho_{2n+1}$, behaves approximately as
\begin{equation}
  \label{eq:asymptotic-odd-ring}
  R_l(\rho) \approx c_{2n+1} \rho^{-1/2} 
  \cos[n_1 k_0(\rho - \rho_{2n})],
\end{equation}
and in the $(2n+2)$th ring as
\begin{equation}
  \label{eq:asymptotic-even-ring}
  R_l(\rho) \approx c_{2n+2} \rho^{-1/2} 
  \sin[n_0 k_0(\rho - \rho_{2n+1})],
\end{equation}
where $c_{2n+1}$ and $c_{2n+2}$ are constant coefficients. Since the rings are supposed to be a quarter-wavelength thick, the expressions $n_1 k_0 (\rho_{2n+1} - \rho_{2n})$ and $n_0 k_0 (\rho_{2n+2} - \rho_{2n+1})$ are approximately equal to $\frac\pi2$ and the successive extremal values of $R_l(\rho)$ are
\begin{subequations}
  \label{eq:extrema}
  \begin{align}
    R_l(\rho_{2n}) &\approx c_{2n+1} \,\rho_{2n}^{-1/2}, \\
    R_l(\rho_{2n+2}) &\approx c_{2n+2} \,\rho_{2n+2}^{-1/2} \approx
    c_{2n+2} \biggl[\rho_{2n} + \frac{\pi}{2k_0}
    \biggl(\frac1{n_0} + \frac1{n_1}\biggr)\biggr]^{-1/2}.
  \end{align}
\end{subequations}
From the requirement of continuity of $E_\phi \propto R_l'(\rho) / \epsilon(\rho)$ at $\rho = \rho_{2n+1}$ it follows that $c_{2n+2}/c_{2n+1} = -n_0/n_1$. Hence,
\begin{equation}
  \label{eq:condition-on-e-phi}
  \biggl|\frac{R_l(\rho_{2n+2})}{R_l(\rho_{2n})}\biggr| \approx
  \frac{n_0}{n_1} \biggl[1 + \frac{\pi}{2k_0\rho_{2n}}
  \biggl(\frac1{n_0} + \frac1{n_1}\biggr)\biggr]^{-1/2}
  \xrightarrow{n\to\infty}\frac{n_0}{n_1}.
\end{equation}
Thus, a localized mode is obtained only if the refractive index of the central rod, $n_0$, is chosen smaller than that of the first ring, $n_1$. 

Figure~\ref{fig:circular-cavity}(a) shows the geometry of an example cavity constructed with the above algorithm. The system consists of an alternating sequence of rings made of air (refractive index $n_0 = 1$) and a MO material with refractive index $n_1 = 2.25$. This value was chosen as the effective index of the fundamental $s$-polarized guided mode of the heterostructure composed of an air-covered 340-nm-thick layer of BIG (refractive index 2.51) grown on a galium-gadolinium-garnet (GGG, refractive index 1.97) substrate, at wavelength $\lambda = 1.3$\,\textmu m. The radii of the first seven rings are listed in Table~\ref{tab:rings-n-2.25-l-1}. This cavity supports a pair of degenerate localized eigenmodes of azimuthal order $l = 1$. The magnetic field of the even mode is plotted in Fig.~\ref{fig:circular-cavity}(a); the field of the odd mode can be obtained by rotating the map from Fig.~\ref{fig:circular-cavity}(a) by $90^\circ$ in the counter-clockwise direction. In Fig.~\ref{fig:circular-cavity}(b), the radial dependence of these magnetic fields, $R_l(\rho )$, is juxtaposed with that of $\vers z \cdot \vect E \tsub e^* \times \vect E\tsub o$, calculated from Eq.~(\ref{eq:circular-cross-product}). Clearly, $R_l(\rho ) R_l'(\rho )$ is negative throughout the MO rings and positive elsewhere; as a result, no cancellations in the upper integral in Eq.~(\ref{eq:circular-V}) occur. The reduced coupling strength of this pair of modes is $\tilde v = 0.00874$. This is more than an order of magnitude stronger than the corresponding value for the cavity proposed in Ref.~\onlinecite{WangOL05} if it were placed in a uniform magnetic field: $\tilde v = 0.0006$. 

The plot in Fig.~\ref{fig:circular-cavity}(b) shows that $\vers z \cdot \vect E \tsub e^* \times \vect E\tsub o$ takes much larger values in the even-numbered (low-index) rings than in the odd-numbered (high-index) ones. This is because this cross product, given by Eq.~\eqref{eq:circular-cross-product}, is inversely proportional to $\epsilon^2$. A potential way to increase the reduced coupling strength $\tilde v$ still further consists therefore in placing the MO material in the \emph{even-numbered} rings. Of course, owing to the constraint~\eqref{eq:condition-on-e-phi}, the MO material would in this case need to become the lower-index component of the cavity. For systems based on garnets, this approach might not be practical, since it would be necessary to fill the (thin!) slits between MO rings with a material of refractive index even higher than ${\sim}2.5$. On the other hand, this approach would be very well suited to systems based on low-index MO polymers, such as those analysed recently by Jalas \textit{et al.}~\cite{JalasOL2010}.

\subsection{Design of ring-cavity circulators: introduction}
\label{ssec:circulator-design-intro}

The next step in the circulator design consists in coupling the MO ring cavity with waveguides. Side coupling, which is often used for classical ring resonators (consisting of only one ring)~\cite{YarivEL00}, is not applicable in our case. As Fig.~\ref{fig:circular-cavity}(a) shows, far from the centre of the cavity, the field of the cavity mode varies very slowly in the azimuthal direction, so it could only be phase-matched by a waveguide mode with an effective index much lower than the refractive index of air --- which cannot be obtained in practice. It is therefore preferable to transfer energy between the cavity and the waveguides by means of butt-coupling.

Figure \ref{fig:geometry-general} shows the general geometry of the class of structures we have considered. They consist of a resonant cavity, composed of $n\tsub f$~full and $n\tsub s$~split rings with inner and outer radii determined by the procedure described in the previous subsection, and three radially oriented identical waveguides of width $d\tsub{wg}$. The distance from the centre of the cavity to the ends of the waveguides is $\rho\tsub{wg}$, while the widths of the slits in the split rings are denoted by $d_{\phi n}$ with $n = n\tsub f + 1$, $n\tsub f + 2$, $\dotsc$, $n\tsub f + n\tsub s$. The rings and waveguides are made of a MO material with permittivity
\begin{equation}
  \label{eq:rib:permittivity}
  \mat\epsilon\tsub h = 
  \begin{bmatrix}
    \epsilon\tsub h & \I g & 0\\
    -\I g & \epsilon\tsub h & 0\\
    0 & 0 & \epsilon\tsub h
  \end{bmatrix}
\end{equation}
and are embedded in an isotropic medium with permittivity $\epsilon\tsub l$. 

In our simulations, the permittivities $\epsilon\tsub h$ and $\epsilon\tsub l$ were chosen as in the previous subsection, i.e.\ as $\epsilon\tsub h = (2.25)^2$ and $\epsilon\tsub l = 1$. A Faraday rotation of $10$--$20^\circ/\text{\textmu m}$ has been measured in BIG at $\lambda = 620$\,nm~\cite{VertruyenPRB08}. The tail of the Faraday spectrum in this reference approaches a couple of deg/\textmu m, which corresponds to $g \sim 0.05$--$0.1$, close to the near-infrared part of the spectrum. To assess the maximum performance that can in principle be achieved, we fixed $g = 0.1$.

The level of coupling between the cavity and waveguide modes, and hence the circulator's performance, depends of course on the values of all the geometrical parameters, which should therefore be optimized. The space spanned by them is rather large, and it is not possible to scan it exhaustively. Therefore our optimization of the presented structure has been somewhat heuristic. The waveguide width $d\tsub{wg}$ was fixed to 250\,nm. The radii of the MO rings, listed in table \ref{tab:rings-n-2.25-l-1}, were determined with the procedure described in the previous subsection to ensure the existence of a pair of cavity modes with the azimuthal order~$l = 1$ at the wavelength $\lambda = 1300$\,nm. 

With the chosen value of~$g$, the relative frequency splitting of these modes is $\Delta\omega/\omega_0 = 2 g \tilde v = 0.00175$. From Eq.~\eqref{eq:loss:gamma}, in the absence of losses the optimum value of $\gamma/\omega_0$ is 0.00151, i.e., the quality factor $Q \equiv \omega_0 / (2\gamma)$ describing the cavity-waveguide coupling should be $Q = 330$. The quality factors of the chosen cavity with 3 and 4 rings are 163 and 829, respectively~\footnote{All the quality factors mentioned in this and the subsequent paragraph were found by searching numerically for non-trivial solutions of the homogeneous system of equations obtained by imposing the conditions of continuity of $H_z$ and $E_\phi$ on the boundaries of the rings.}. Therefore one can expect that the waveguides of an optimally designed circulator should end somewhere close to the third innermost ring --- or possibly even inside it, since the coupling to waveguides is doubtlessly less efficient than that to the whole surrounding free space.

The total number of rings necessary for ensuring a prescribed level of peak transmission $T\tsub{min}$ could in principle be estimated from Eq.~\eqref{eq:loss:min-Q-l}: for instance, for $T\tsub{min} = 0.9$ the quality factor~$Q\tsub r$ describing radiation loss should be greater than 6430. This is already ensured by a 6-ring cavity, whose quality factor reaches 21,140. However, the quality factors of cavities with split outer rings will necessarily be smaller than of those with full rings; therefore, a larger number of rings might be necessary to ensure a 90-percent efficiency. In our calculations, we restricted our attention to systems with at most 7~rings. 

\begin{table}
  \centering
  \begin{tabular}{ccc}
    \toprule
    Ring number & Inner radius (nm) & Outer radius (nm) \\
    \midrule    
    1 & \hphantom{0}381 & \hphantom{0}539\\
    2 & \hphantom{0}847 & \hphantom{0}998\\
    3 & 1309 & 1457\\
    4 & 1772 & 1919\\
    5 & 2236 & 2382\\
    6 & 2700 & 2846\\
    7 & 3165 & 3310\\
    \bottomrule
  \end{tabular}
  \caption{Radii of the high-index rings of the cavity with $\epsilon\tsub h = (2.25)^2$ and $\epsilon\tsub l = 1$ supporting a resonant mode with azimuthal order $l = 1$ at wavelength $\lambda = 1300$\,nm.}
  \label{tab:rings-n-2.25-l-1}
\end{table}

Having fixed the number of split and full rings of the circulator in this way, we were essentially left with the problem of optimizing the values of $d_{\phi n}$ and $\rho\tsub{wg}$. The ultimate figure of merit, the bandwidth, is nonzero only for structures already rather close to optimum, which makes it a cumbersome objective function. Therefore we optimized instead the maximum isolation level, hoping that circulators with large values of this parameter would also be characterized by a large bandwidth.

\begin{figure}
  \centering
  \includegraphics{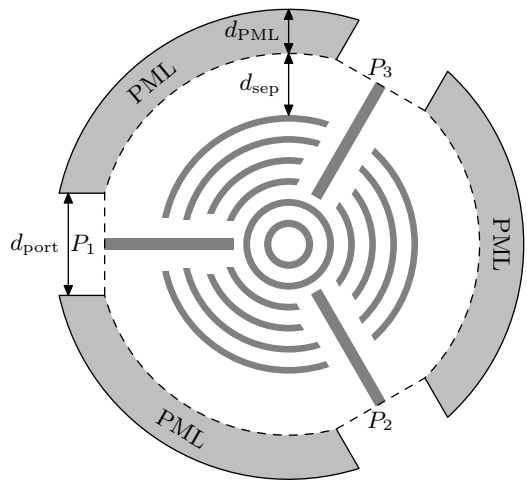}
  \caption{Geometry of the domain used in FEM calculations.}
  \label{fig:rib:computational-domain}
\end{figure}

Numerical calculations of the transmission through the circulators studied in this section were performed with the finite-element method (FEM) using the RF module of the COMSOL software package. To account for the presence of idealized semi-infinite waveguides, the computational domain, shown in Fig.\ \ref{fig:rib:computational-domain}, was constructed as follows. The region surrounded by the dashed line is a fragment of the physical system shown in Fig.\ \ref{fig:geometry-general}. The parts of its boundary lying ``far'' from the waveguide ends are covered with perfectly-matched layers (PMLs) implemented by means of the complex coordinate transform $\rho \mapsto s\tsub{PML}(1+\I)(\rho - \rho\tsub{in}) / d\tsub{PML}$, where $\rho$~is the radial coordinate measured form the centre of the cavity, $\rho\tsub{in}$ denotes the radius of the inner PML boundary, $d\tsub{PML} = 600$\,nm is the PML thickness, and $s\tsub{PML} = \lambda$ its strength. The radius $\rho\tsub{in}$ is chosen so that the distance from the outermost ring to the inner PML boundary is $d\tsub{sep} = 5000$\,nm. Perfect-electric-conductor boundary conditions are imposed on the parts of the domain's boundary adjacent to the PML. In constrast, the electromagnetic fields on the segments $P_n$ ($n = 1, 2, 3$) of length $d\tsub{port} = 2250$\,nm, perpendicular to the waveguides, are constrained to be a superposition of the incoming and outgoing guided modes of the corresponding waveguides. The profile of these modes is calculated analytically and normalized to unitary power, and the amplitude of the incoming mode is set to unity on~$P_1$ and to zero on $P_2$ and~$P_3$. Physically, these constraints correspond to the assumption that all the radiative waveguide modes excited by the cavity decay before reaching the ports~$P_n$. The domain is divided into triangular nodal Lagrangian elements of order~$p = 5$ and maximum allowed size $h\tsub{max}/\sqrt\epsilon$, where $h\tsub{max} = 500$\,nm and $\epsilon$ is the local permittivity.

We have verified that above choice of the computational parameters allows to calculate the 20-dB-bandwidth of typical circulators with an accuracy of about $3\%$ (see Ref.\ \onlinecite{WojtekPhD}, Section 4.5.2, for a detailed convergence study).

\subsection{Design of ring-cavity circulators: geometry optimization}
\label{ssec:circulator-design-optimization}

\begin{figure}[htb!]
  \centering
  \includegraphics{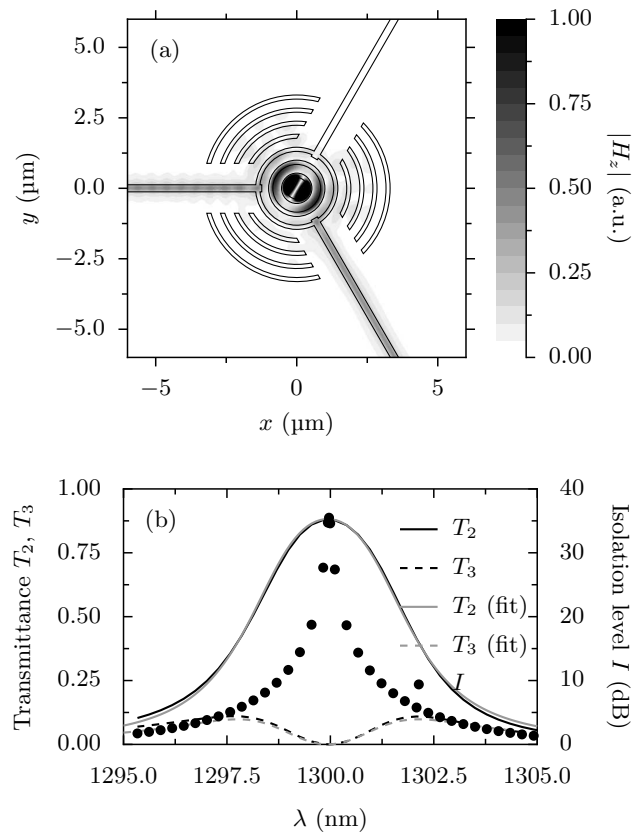}  
  \caption{\subnumber{(a)} Magnitude of the magnetic field in circulator~\ref{circ:best-rectangular} at the wavelength $\lambda = 1300.0$\,nm, corresponding to the maximum isolation level. The waveguide mode is incident from the left. \subnumber{(b)}~Wavelength dependence of the transmission (left axis) and isolation level (right axis) of~\ref{circ:best-rectangular}.}
  \label{fig:rib:C22}
\end{figure}

We initially focused on circulators with $d_{\phi n}$ independent from~$n$, i.e., with waveguides enclosed in rectangular ``slits'' of width $d_\phi$. The contours in Figure \ref{fig:rib:C22}(a) show the geometry of the best structure we have found, called~\ref{circ:best-rectangular} in the following. It has 3~full and 4~split rings. The slit width $d_{\phi} = 1770$\,nm and the waveguide ends are located at $\rho\tsub{wg} = 1240$\,nm from the cavity centre, so that the waveguides cross the outermost full ring. For ease of reference, the values of all the geometrical parameters of~\ref{circ:best-rectangular} are listed in table \ref{tab:circ-parameters}. This circulator offers maximum isolation level of 35\,dB, and the wavelength dependences of $T_2$, $T_3$ and $I$ are shown in Fig.\ \ref{fig:rib:C22}(b). Clearly, the curves are fairly symmetric with respect to the central wavelength $\lambda = 1300.0$\,nm, which indicates that the direct coupling between waveguides is insignificant. Figure \ref{fig:rib:C22}(c) shows the map of the magnetic field at $\lambda = 1300.0$\,nm. At this wavelength, 88\% of the input power is transmitted to waveguide~2; the rest is mainly lost to the surrounding free space. Far from the peak, the amount of these losses can exceed 50\%. This behaviour contrasts with that of the PC circulator, where almost 100\% of the input power remains in the waveguide system due to the quasi-perfect isolation provided by the surrounding periodic lattice. The 20-dB bandwidth $B(100)$ of circulator~\ref{circ:best-rectangular} is 0.729\,nm (129\,GHz).

\begin{table}[!b]
  \centering
  \begin{tabular}{ccccccc}
    \toprule
    Circulator &
    $n\tsub f$ &
    $n\tsub s$ &
    $\rho\tsub{wg}$ (nm) &
    $d_\phi$ (nm) & 
    $\phi\tsub{cone}$ ($^\circ$) & Slits\\
    \midrule
    \ref{circ:best-rectangular} & 3 & 4 & 1240 & 1770 & --- & rectangular\\
    \ref{circ:best-conical} & 3 & 4 & 1210 & ---& 35.1 & conical \\
    \bottomrule
  \end{tabular}
  \caption{Geometrical parameters of the circulators analysed in the text. The ring radii are listed in the first $(n\tsub f + n\tsub s)$ rows of table~\ref{tab:rings-n-2.25-l-1}.}
  \label{tab:circ-parameters}
\end{table}

The grey curves in Fig.\ \ref{fig:rib:C22}(b) show the transmittance curves of this circulator as predicted by the coupled-wave model with radiation losses taken into account, presented in Section~\ref{ssec:tcm_radiation}. The values of the parameters $\omega_0$, $\Delta\omega$, $\gamma$ and $\gamma\tsub r$ were found by fitting the expressions for $T_2$ and $T_3$ obtained from eqs.~\eqref{eq:loss:scattering-coeffs} to the values calculated numerically. The Levenberg-Marquardt algorithm was used as the fitting procedure. The best fit was obtained for parameters corresponding to $\lambda_0 \equiv 2\pi c / \omega_0 = 1299.9$\,nm, $\Delta \lambda \equiv 2\pi c \Delta\omega / \omega_0^2 = 2.3$\,nm, $Q \equiv \omega_0 / (2\gamma) =370$ and $Q\tsub r \equiv \omega_0 / (2\gamma\tsub r) = 5730$. Clearly, there is a good match between the theoretical and numerical curves; its quality would probably be further improved by taking into account the  direct coupling between waveguides, which causes the slight asymmetry of the numerical plots. The quality factor related to losses, $Q\tsub r$, is significantly lower than that of an isolated cavity with 7 full rings, which is as large as 107,000. This is obviously due to the presence of slits. On the other hand, the position of the ends of the waveguides (just inside the third innermost ring) is in good accord with the predictions made in Section~\ref{ssec:circulator-design-intro}.

\begin{figure}
  \centering
  \includegraphics{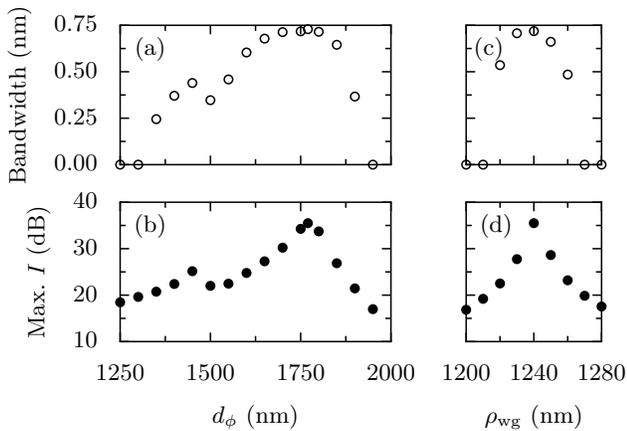}
  \caption{Tolerance of the bandwidth $B(20\,\text{dB})$ and the maximum isolation level $I\tsub{max}$ of circulator~\ref{circ:best-rectangular} to perturbations of the parameters \subnumber{(a)--(b)}~$d_\phi$ and \subnumber{(c)--(d)}~$\rho\tsub{wg}$.}
  \label{fig:rib:C22-perturbations}
\end{figure}

We have found this device very tolerant to variations of the slit width $d_{\phi}$; Figs.\ \ref{fig:rib:C22-perturbations}(a)--(b) show the dependence of the maximum isolation level and bandwidth of~\ref{circ:best-rectangular} on this parameter. It can be seen that the bandwidth stays above 0.5\,nm in a 300-nm-wide range  of $d_{\phi}$. The constraints on $\rho\tsub{wg}$ are more stringent: as shown in Fig. \ref{fig:rib:C22-perturbations}(c), the corresponding range of $\rho\tsub{wg}$ is about 40-nm wide.

The performance of this structure changes rather abruptly when the number of rings is modified. For instance, if the seventh ring is removed, the maximum isolation level decreases to only 21\,dB and the bandwidth to 0.386\,nm (69\,GHz). However, it is possible to improve these figures by readjusting the slit widths and the position of the waveguide ends: for $\rho\tsub{wg} = 1250$\,nm and $d_{\phi} = 1570$\,nm (10\% less than in the 7-ring case) $I\tsub{max}$ reaches 33\,dB and the bandwidth 0.603\,nm (107\,GHz). 

\begin{figure}
  \centering
  \includegraphics{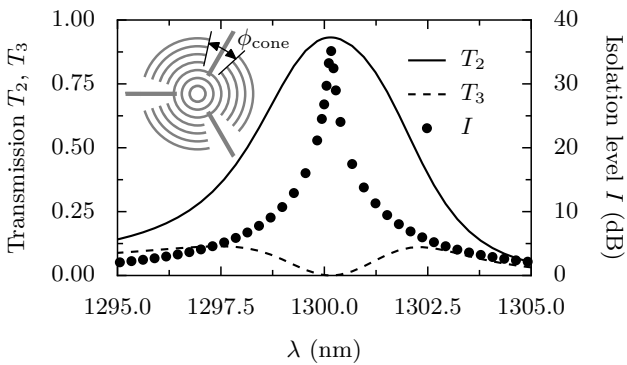}

  \caption{Wavelength dependence of the transmission (left axis) and isolation level (right axis) of circulator~\ref{circ:best-conical} (inset) with waveguides enclosed in conical slits.}
  \label{fig:rib:conical}
\end{figure}

Having noted that both for 6 and 7~rings the optimum angular length of the removed sectors of the outermost ring is almost the same, ${\sim}34^\circ$, we test the performance of a second class of structures, in which the outer rings are truncated along radial lines instead of ones parallel to the waveguides. The latter are thus enclosed by conical rather than rectangular air slits, as illustrated in the inset of Fig.\ \ref{fig:rib:conical}. We found the optimum cone aperture $\phi\tsub{cone}$ to be $35.1^\circ$, close to the value cited above. The optimum position of the waveguide end, $\rho\tsub{wg} = 1210$\,nm, is also only slightly different from the original one. More importantly, the maximum isolation level and bandwidth decrease much less (from 35 to 28\,dB and from 0.770 to 0.708\,nm, or 137 to 126\,GHz, respectively) when the seventh ring is removed. This relative insensitivity to the details of the geometrical structure of the exterior region of the device is the behaviour that one would intuitively expect from a well-designed circulator; therefore, structures with conical slits seem closer to the ideal than those with rectangular ones. Figure \ref{fig:rib:conical} shows the wavelength dependence of $T_2$, $T_3$ and~$I$ of the optimum 7-ring circulator with conical slits, referred to as~\ref{circ:best-conical}. Its parameters are listed in table \ref{tab:circ-parameters}.

\section{Perspectives --- towards 3D nonreciprocal cavities}\label{sec:perspectives}

\declaregeometry{geom:si}
\declaregeometry{geom:sym}
\declaregeometry{geom:asym}

\subsection{High-contrast 3D cavities}\label{ssec:high3D}
It is well known that the properties of a 3D cavity depend crucially on its vertical multilayer profile, and a cavity with a high $Q$ in 2D can leak heavily in 3D if the presence of the substrate and superstrate is not taken into account. The design algorithm of Section~\ref{ssec:cavity_design} is based on 2D considerations, and a similar algorithm in 3D is nearly impossible to develop. However, the 2D algorithm has allowed us to compare our designs to previously reported circulator cavities~\cite{WangAPB05,WangOL05,WangPhotNano06}. So far, the true 3D nature of a real integrated system has only been taken into account in our work by using an effective-index approximation for the unetched MO rings. The material between the rings has, however, always been set to be air --- an approach that is reasonable only if the garnet layer is entirely etched away between the MO rings. Such a high-contrast ring cavity cannot, however, be fabricated in a membrane form, since the rings are not interconnected. Therefore, in practice the rings will need to be placed on a solid substrate with refractive index~$n\tsub{sub}$. Far from its centre, the ring cavity can be locally approximated by a linear grating with period
\begin{equation}
\Lambda = \frac{\lambda}{4}\biggl(\frac{1}{n_0}+\frac{1}{n_1}\biggr).
\label{eq:Bragglinear}
\end{equation}
If $\Lambda > \lambda/2n_\mathrm{sub}$, all Bloch eigenmodes of the grating will leak into the substrate. In order to avoid that, $n_\mathrm{sub}$ must be smaller than $\lambda/2\Lambda$. For $n_0=1$, $n_1=2.25$ and $\lambda=1300\,\text{nm}$, the substrate index must be less than 1.4. This might prove very difficult to achieve.

\subsection{Low-contrast 3D cavities}\label{ssec:low3D}
In low-index-contrast cavities substrate leakage is much easier to avoid, albeit at the price of larger cavity due to smaller Bragg gap. This can be seen by rewriting the no-leakage condition as $n_\mathrm{sub}/2 < n_{\mathrm{eff},0}n_{\mathrm{eff},1}/(n_{\mathrm{eff},0}+n_{\mathrm{eff},1})$. This condition is always met if both effective indices are larger than the substrate index $n\tsub{sub}$, around 1.9 for typical garnet substrates (such as GGG).

\begin{figure}[!tb]%
\centering
\includegraphics{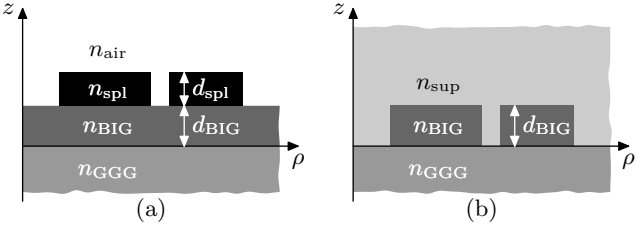}%
\caption{Schematics of 3D structures of types \subnumber{(a)} \ref{geom:si} and \subnumber{(b)} \ref{geom:sym}.}%
\label{fig:3Dlayouts}%
\end{figure}

We have considered two classes of low-contrast cavities. In one of them, shown in Fig.\ \ref{fig:3Dlayouts}(a) and labelled G1 henceforth, the garnet layer ($d_\mathrm{BIG}=280$\,nm) is left unetched and the cavity is formed by etching in a high-index cladding layer ($n\tsub{spl}=3.5$, $d\tsub{spl}=80$\,nm). The MO properties of the exposed BIG regions are assumed to be destroyed by e.g.\ ion implantation. In the second class of cavities, shown in Fig.\ \ref{fig:3Dlayouts}(b) and referred to as \ref{geom:sym}, rings etched in BIG ($d\tsub{BIG} = 330$\,nm) are buried in a material with index $n\tsub{sup}$ approximately equal to that of the GGG substrate. A potential candidate is silicon nitride (Si$_3$N$_4$); around $\lambda = 1300$\,nm, the refractive index of both GGG and Si$_3$N$_4$ is 1.97. As indicated above, the use of the effective-index approximation to determine the position and width of the rings by help of the algorithm of Section~\ref{ssec:cavity_design} is now better justifiable. We have used an in-house-developed rigorous FEM code \cite[Section 5.3]{WojtekPhD} to compute the resonant frequencies of axisymmetric 3D cavities belonging to the just described classes, crosschecking the data obtained with the 2D model. Table~\ref{tab:mph:3d} summarizes the results of these calculations.

\begin{table}
  \centering\floatfont
  \begin{tabular}{l*{5}{>{$}c<{$}}}
    \toprule
    Type & l & \Delta\omega/\omega_0$ (3D)$ & \Delta\omega/\omega_0$ (2D)$ & Q$ (3D)$ & Q$ (2D)$ \\ 
    \midrule
    \ref{geom:si}  & \phz1  & 0.00129   & 0.00166  & 104 & 1277\\ 
    & \phz4  & 0.00200  & 0.00286  & 105 & 1410 \\ 
    & 10 & 0.00227  & 0.00342 & 109 & 1360 \\ 
    \midrule
    \ref{geom:sym}  & \phz1  & 0.00138   & 0.00207 & 134 & 2079 \\ 
    & \phz4  & 0.00213  & 0.00349  & 172 & 2285 \\ 
    & 10 & 0.00241  & 0.00412 & 334 & 2139 \\ 
    \midrule
    \ref{geom:sym}$^a$ & 10  & 0.00261   & $---$ & 2420 & $---$ \\ 
    \bottomrule
  \end{tabular}
  
  \smallskip
  {\small $^a$ After numerical optimization of ring radii.}
  \caption{Relative frequency splitting $\Delta\omega/\omega_0$ and quality factor $Q$ of the eigenmodes of several \ref{geom:si}- and \ref{geom:sym}-type 20-ring cavities designed to support modes of azimuthal order~$\pm l$. The values obtained with rigorous 3D FEM calculations are compared to the estimates obtained with 2D calculations in the framework of the effective-index approximation. Note that $Q$ is defined as the average of the quality factors of the modes with aximuthal order $l$ and~$-l$.}
  \label{tab:mph:3d}
\end{table}

They prove first of all that the 2D design procedure is flexible enough to let us adapt it to 3D layouts. The frequency splittings calculated with the FEM code are only slightly reduced with respect to those predicted by the 2D model. In fact this decrease can be entirely attributed to a reduced mode confinement in the MO layer due to the vertical permittivity profile. The more worrying aspect, however, is the strong diminution of the quality factor. The quality factors of the 3D cavities are an order of magnitude smaller than those obtained with 2D calculations (for the same number of rings and related only to the in-plane confinement). 

The negative impact of large radiation losses (and thus low $Q$ factors) on the operation of the nonreciprocal cavity can be easily understood by considering the following equations of the coupled-mode model including radiation losses:
\begin{subequations}
\label{eq:lowQ}
\begin{align}
\label{eq:3Dtransmission}
    s\tsub{3,out} &= 
    -\frac23
    \frac{\gamma [\gamma + \gamma\tsub r - \tilde V \sqrt 3 -
      \I(\omega-\omega_0)]}
    {[\gamma +\gamma\tsub r - \I(\omega-\omega_0)]^2 + \tilde V^2},\\
\label{eq:optimumQ3D}
    \frac{1}{Q} + \frac{1}{Q \tsub r} &= \frac{\Delta\omega}{\omega_0} \sqrt 3,
\end{align}
\end{subequations}
where $Q$ and $Q\tsub r$ describe respectively the ``useful'' cavity-waveguide coupling and the radiation loss. The latter equation is obtained by rewriting Eq.~\eqref{eq:loss:gamma} using $2\abs{\tilde V} = \Delta\omega$ and the definition of the quality factor $Q\equiv\frac{\omega_0}{2\gamma}$. For simplicity only the case without direct coupling ($\Delta=\frac12\pi$) is considered. Equation~\eqref{eq:optimumQ3D} expresses the optimum relationship between the relative frequency splitting and the two $Q$ factors. It implies that for a radiative quality factor $Q \tsub r$ lower than  $\omega_0/(\Delta\omega\sqrt 3)$ perfect circulation cannot occur, since it would require a negative (unphysical) value of~$Q$. For such small $Q\tsub r$ the coupling necessarily becomes overcritical, i.e.\ $\gamma \tsub r + \gamma$ is larger than the optimum value. 

Fig.~\ref{fig:isolation-maps}(a) shows that the circulator has some tolerance with respect to $\gamma$. At a ten-percent overtuning ($\gamma \sim 2\abs{\tilde V}$) maximum isolation level of ${\sim}20$\,dB is still possible. This will of course come at the price of a heavily reduced transmission level. Indeed, assuming a modest overtuning (in order to keep the isolation level reasonable), from Eq.~\eqref{eq:3Dtransmission} we obtain $\max T_3 \equiv \max \abs{s\tsub{3,out}}^2 \sim (\frac{\gamma}{\sqrt{3}\tilde V})^2$, with $\gamma = \frac{\omega_0}{2Q}$ describing the cavity-waveguide coupling alone. Thus, a ten-percent overtuning with $Q\tsub r \sim \omega_0/(\Delta\omega\sqrt{3})$ and $Q \sim 0.1\omega_0/(\Delta\omega\sqrt{3})$ would lead to transmission levels of about 1\%.

\subsection{Towards low-contrast nonreciprocal cavities with higher quality factor}
Table~\ref{tab:mph:3d} shows that none of the 3D cavities under consideration supports a dipolar ($l = 1$) mode with an inverse radiation quality factor $1/Q\tsub r$ smaller than 400\% of the optimum coupling strength, $\Delta\omega\sqrt{3}/\omega_0$. The most direct way to increase~$Q\tsub r$ consists in moving towards designs for higher azimuthal orders $l$. The stronger azimuthal variation leads to higher in-plane $k$-components in the cavity mode and thus lower substrate leakage. Conveniently, according to~\eqref{eq:circular-V}, the relative frequency splitting also increases with azimuthal order. The price to pay is a higher modal volume. Table~\ref{tab:mph:3d} shows nevertheless that the increase in $Q\tsub r$ is limited. The best cavity designed for $l=10$, the one belonging to the class \ref{geom:sym}, supports a mode with $1/Q\tsub r$ of about 70\% of the optimum coupling strength $\Delta\omega\sqrt3/\omega_0$. As a result, at optimum coupling, $\gamma = \frac{\omega_0}{2}\frac{1}{Q} \sim \frac{\omega_0}{2} 0.3 \frac{\Delta\omega}{\omega_0}\sqrt{3}=0.3\sqrt{3}\abs{\tilde V}$ and one still cannot expect transmission levels larger than 10\% in the best-case scenario.

\begin{figure}[!htb]
  \centering\floatfont
  \includegraphics{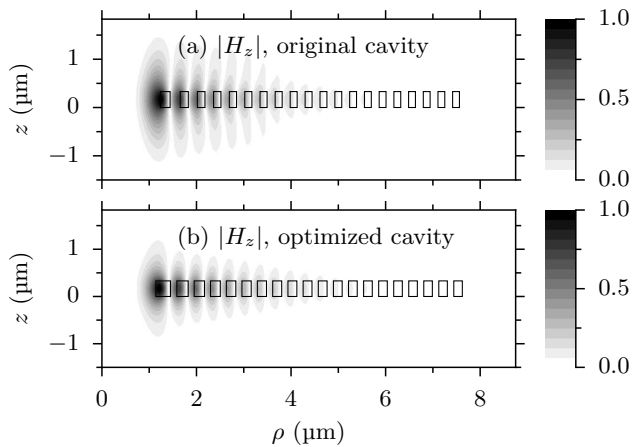}
  \caption{Magnitude of the $z$ component of the magnetic field of the eigenmode of \subnumber{(a)} the original \ref{geom:sym}-type cavity designed for $l = 10$, and of \subnumber{(b)} its numerically optimized version; both in the absence of magnetization. The rectangles denote the cross-sections of BIG rings immersed in a material with refractive index 1.97. The improved field confinement (and thus higher~$Q$) is easily seen. The MO interaction described by the cross product $\vect E\tsub e^* \times \vect E\tsub o$ (not shown here) is almost not influenced.}
  \label{fig:mph:opt:geometries}
\end{figure}

We have recently applied a numerical optimization procedure to adjust the positions and widths of the rings. In a first attempt this has already led to an improved design for a 20-ring \ref{geom:sym}-type cavity (for an $l=10$ mode). Fig.~\ref{fig:mph:opt:geometries} shows the geometry of the original and the optimized \ref{geom:sym}-type cavity, as well as the maps of the normalized magnetic field of their modes. The $Q$ factor of the mode of the optimized cavity is as large as 2420: an increase by almost an order of magnitude from 334. Moreover, it is higher than that of the mode of the original 2D 20-ring cavity, 2139. This means that the $Q$ factor of the optimized cavity is constrained by in-plane rather than out-of-plane radiation loss, and so it could be further augmented if necessary by increasing the number of rings. At the same time, the relative frequency splitting stays quite tolerant with respect to this optimization, indicating that as far as MO effect goes the 2D design algorithm proposed here is already quasi-optimal. The obtained $\Delta\omega/\omega_0$ is 0.00261, even slightly larger than in the original structure. Using similar estimations as above, these values imply that in a best-case scenario perfect circulation with a maximal transmission close to 90\% is possible. The total device footprint remains close to $(10\lambda)^2$. This work together with a 3D analysis of the waveguide-cavity coupling problem will be the subject of a future communication.

\section{Conclusion}
In this work we have presented an extensive study of a new concept for a very compact and technologically feasible type of integrated optical circulator. Its operation is based on a MO frequency splitting of a pair of degenerate cavity modes of opposite (azimuthal) parity. In contrast to previous reports on MO cavity circulators, the type proposed and studied here operates without oppositely magnetized domains or a full $k$-space bandgap. The confinement is provided by a radial Bragg mirror consisting of uniformly magnetized, concentric MO rings. An original design algorithm taking into account the analytical properties of the functions describing the radial profile of the cavity modes allows to determine the optimum position and width of the rings. We have extended an abstract coupled-mode model to verify the impact of radiation losses and direct transmissions between the circulator ports. These might arise due to the absence of a full bandgap isolating the cavity ports.

The extended coupled-mode model has indicated that the presence of direct pathway transmissions leads unexpectedly to interesting alternative operation regimes of these nonreciprocal cavities. The combination of nonreciprocity and Fano-type interactions allows to achieve circulation regimes where the sign of the circulation can be spectrally tuned. Howsoever interesting the regime at high direct coupling is, devices designed to operate without direct coupling are expected to be less sensitive to perturbations of the cavity-port coupling level from its optimum value. We have demonstrated numerically the satisfactory circulator operation of a 6- and 7-ring cavity partially intersected by three 120$^\circ$-spaced rib waveguides. Circulation extinction as high as 35\,dB, and an insertion loss below 1\,dB over a 20\,dB circulation bandwidth of 130\,GHz (around 1300\,nm), have been demonstrated. The device is found to have tolerances on the position and dimensions of its features that are well within the limits of modern nanotechnology.

Looking forward to realistic 3D device layouts, we have studied different types of high- and low-contrast cavities. An FEM tool for the calculation of the eigenmodes of axisymmetric, 3D MO cavities has been developed for that purpose. Air-suspended high-contrast ring cavities being physically impossible, the minimum requirements for the $Q$-factor of the unavoidable low-contrast cavities have been indicated. First attempts at geometry optimization of such 3D cavities prove convincingly that the quality factor of low-contrast ring cavities can be raised enough to obtain devices with operation characteristics similar to those achieved in 2D designs. Theoretical and experimental work is ongoing on the optimization and demonstration of such an ultracompact integrated optical circulator.

\section*{Acknowledgments}
The authors acknowledge financial support of the French National Research Agency (ANR) via funding through the ANR-MAGNETOPHOT program.
\bibliographystyle{model1-num-names-with-eid}
\bibliography{mybibref}

\end{document}